\documentclass[12pt]{article}
\usepackage{amsfonts,graphicx,amsmath,amsthm,amssymb,epsfig}
\usepackage{float}
\allowdisplaybreaks
\begin{document}

\title{\bf Study of Charged Cylindrical Collapse in $f(\mathcal{R},\mathcal{T},\mathcal{Q})$ Gravity}
\author{M. Sharif~$^1$ \thanks{msharif.math@pu.edu.pk}~ and Tayyab Naseer~$^{1,2}$ \thanks{tayyabnaseer48@yahoo.com;
tayyab.naseer@math.uol.edu.pk}\\
$^1$ Department of Mathematics and Statistics, The University of Lahore,\\
1-KM Defence Road Lahore, Pakistan.\\
$^2$ Department of Mathematics, University of the Punjab,\\
Quaid-i-Azam Campus, Lahore-54590, Pakistan.}

\date{}
\maketitle

\begin{abstract}
This paper investigates the effects of electromagnetic field on the
gravitational collapse in $f(\mathcal{R},\mathcal{T},\mathcal{Q})$
theory, where
$\mathcal{Q}=\mathcal{R}_{\varphi\vartheta}\mathcal{T}^{\varphi\vartheta}$.
For this, we assume dynamical cylindrically symmetric
self-gravitating geometry which is coupled with generalized
anisotropic matter distribution as well as dissipation flux. We
adopt the model $\mathcal{R}+\Phi\sqrt{\mathcal{T}}+\Psi\mathcal{Q}$
to formulate the corresponding dynamical and transport equations by
employing the Misner-Sharp as well as M\"{u}ler-Israel Stewart
formalisms, where $\Phi$ and $\Psi$ are real-valued coupling
constants. The influence of state variables, heat dissipation,
charge and the bulk viscosity on the collapsing phenomenon is then
studied by establishing some relations between these evolution
equations. Moreover, the Weyl scalar and the modified field
equations are expressed in terms of each other. We apply some
constraints on the considered modified model and the fluid
configuration to obtain conformally flat spacetime. Finally, we
address different cases to check how the modified corrections and
charge affect the collapse rate of cylindrical matter source.
\end{abstract}

\section{Introduction}

According to cosmological observations, the expansion of superheated
matter and energy is considered as the origin of our universe.
Cosmologists explored that a considerable portion of this
unfathomable universe is composed of stars, planets and galaxies.
The most appealing and promising phenomenon in this regard is the
gravitational collapse due to which the structural formation of
these celestial objects takes place. The pioneer work of
Chandrasekhar \cite{28} on this phenomenon helps scientists to
understand its importance in the field of relativistic astrophysics.
He found that a star remains stable until its external pressure and
internal force of attraction (due to its mass) are counterbalanced
by each other. Oppenheimer and Snyder \cite{29} studied the dynamics
of the dust collapse and found that such collapse ultimately leads
to the formation of a black hole. The anisotropic spherically
symmetric geometry has been discussed by Misner and Sharp
\cite{29a}, through which they checked the role of pressure
anisotropy on the collapse rate. Herrera and Santos \cite{30}
investigated this phenomenon for a dissipative sphere by using the
Misner-Sharp technique. They observed the dissipation of energy from
that compact body by means of heat and radiations. Herrera \emph{et
al.} \cite{31} analyzed the most general cylindrically symmetric
matter configuration and found that the collapse of such an object
eventually produces gravitational waves outside that source.

The influence of electromagnetic field on massive objects plays a
considerable role in studying their evolution and stability. The
effects of magnetic and Coulomb forces help to overcome the
gravitational attraction of a compact object, thus a considerable
amount of charge is required to make that system more stable. The
self-gravitating uncharged/charged structures have been discussed by
Sharif and his associates \cite{32} through different approaches and
it was observed that the presence of charge ultimately reduces the
collapse rate. The gravitational collapse is immensely dissipative
process, thus the effects of dissipation flux in this phenomenon
cannot be ruled out \cite{32a}. Chan \cite{34} examined a radiating
isotropic geometry and revealed that the presence of shear viscosity
makes the fluid more and more anisotropic. They also found that the
collapsing time of the considered star is same in the absence and
presence of shear viscosity. Di Prisco \emph{et al.} \cite{35c}
studied anisotropic charged sphere with shear and discussed the role
of charge on inhomogeneity of the energy density. Nath \emph{et al.}
\cite{35a} employed matching criteria by taking quasi-spherical
Szekeres spacetime as an interior geometry and charged Vaidya metric
as exterior, and examined the collapsing rate. They found that the
electric charge supports the naked singularity to form. Herrera
\emph{et al.} \cite{35} discussed viscous dissipative distribution
and observed that the parameters which govern heat dissipation
lessen the gravitational pull that eventually leads to the reduction
of collapse rate.

The cylindrical symmetry is supported by the existence of
cylindrical gravitational waves whose study produces remarkable
consequences. Several fundamental properties of these geometrical
objects have been discussed in the literature. Bronnikov and
Kovalchuk \cite{35d} pioneered the extensive analysis of such
structures. Wang \cite{35e} considered four-dimensional cylinder
along with a massless scalar field and formulated exact solutions to
the corresponding field equations. They found that black hole may
eventually be the result of gravitational collapse of such object.
Guha and Banerji \cite{36} studied anisotropic charged cylinder
experiencing heat dissipation and investigated the influence of
charge as well as dissipative entities on the collapsing source
through the dynamical equations. A quantity which is of great
importance in the stellar evolution of celestial structure and helps
to measure its curvature is the Weyl tensor. Penrose \cite{36a}
discussed the collapsing rate of a sphere by expressing a particular
relation between physical quantities and the Weyl tensor. Sharif and
Fatima \cite{36b} studied the restricted case of cylindrical
spacetime and demonstrated the relationship between matter
variables, anisotropic pressure, coefficient of shear viscosity and
the Weyl tensor. They confirmed the fulfillment of conformal
flatness condition for the homogeneous energy density, and vice
versa.

An enticing phenomenon in the field of cosmology and astrophysics
for the last few years is the ongoing cosmic accelerated expansion
\cite{1a}. Several astronomers claimed that a mysterious form of
force triggers such expansion due to its abundance in the universe,
named as the dark energy that has immense repulsive effects. The
study of this current evolutionary phase in the context of general
theory of relativity ($\mathbb{GR}$) faces certain short comings
such as fine-tuning and cosmic coincidence problem. In order to find
suitable solution to these issues, scientists developed multiple
extensions to $\mathbb{GR}$. Among them, $f(\mathcal{R})$ theory is
the first and foremost possible generalization, obtained by
replacing the Ricci scalar $\mathcal{R}$ with its generic functional
in the Einstein-Hilbert action \cite{1c}, to study cosmic structures
at large scales. Multiple approaches have been employed in this
framework to construct solutions corresponding to various
$f(\mathcal{R})$ models and the acquired results are found to be
physically acceptable \cite{2}-\cite{2e}.

Initially, Bertolami \emph{et al.} \cite{10} presented the notion of
coupling between matter and geometry to study the appealing nature
as well as composition of the cosmos. They engaged the geometrical
terms (i.e., the Ricci scalar) in the matter Lagrangian and then
studied the influence of such coupling in $f(\mathcal{R})$
framework. A couple of years ago, Harko \emph{et al.} \cite{20}
generalized this interaction at action level by introducing
$f(\mathcal{R},\mathcal{T})$ gravity, in which $\mathcal{T}$ is
trace of the energy-momentum tensor $(\mathbb{EMT})$. Such
gravitational theories provide non-vanishing divergence of their
corresponding $\mathbb{EMT}$, and thus opposing $\mathbb{GR}$ as
well as $f(\mathcal{R})$ gravity. Several self-gravitating
structures have been investigated in this theory from which multiple
surprising results are obtained \cite{21a}-\cite{21f}. This theory
cannot entail the coupling effects on celestial bodies in some
particular situations, thus one needs to overcome this flaw. Haghani
\emph{et al.} \cite{22} further added a factor $\mathcal{Q}$ in this
theory, representing contraction of the Ricci tensor with
$\mathbb{EMT}$, and named this theory as
$f(\mathcal{R},\mathcal{T},\mathcal{Q})$. They considered three
different non-minimally coupled models and discussed their
cosmological applications for different configurations such as high
density regime and pressureless fluid case.

Sharif and Zubair \cite{22a} adopted two modified models
$\mathcal{R}+\lambda\mathcal{Q}$ and
$\mathcal{R}(1+\lambda\mathcal{Q})$, and discussed first and second
laws of the black hole thermodynamics corresponding to two different
choices of the matter Lagrangian. They also addressed energy bounds
within the same context and concluded that weak energy conditions do
not hold for negative values of the coupling constant $\lambda$
\cite{22b}. Odintsov and S\'{a}ez-G\'{o}mez \cite{23} studied the
flat FLRW metric for several cosmological extended models and
studied the problem of fluid instability. The gravitational action
of this theory has been reconstructed and de Sitter universe
solutions for the perfect fluid configuration are also presented by
them. Two particular $f(\mathcal{R},\mathcal{T},\mathcal{Q})$ models
have been discussed by Baffou \emph{et al.} \cite{25}, from which
they concluded that the convergence of matter and geometrical
perturbation functions leads to the stability of this theory.
Recently, we have discussed anisotropic solutions with/without
considering the effects of electromagnetic field and found them
physically acceptable \cite{27,27aa}.

The gravitational collapse of spherical as well as cylindrical
spacetimes has extensively been studied in various modified
theories. Borisov \emph{et al.} \cite{27ab} employed numerical
simulations to study the spherical collapse in the background of
$f(\mathcal{R})$ gravity. Shamir and Fayyaz \cite{27ac} explored a
dynamical cylinder whose interior distribution involves anisotropic
pressure and dissipative flux in the same theory. They found that
the collapse plays a governing role in discussing the late-time
acceleration. Various authors investigated anisotropic
self-gravitating spherical/cylindrical sources and discussed their
collapsing rate through the dynamical equations in
$f(\mathcal{R},\mathcal{T})$ theory \cite{27ad,27add}. Bhatti
\emph{et al.} \cite{27aad} analyzed anisotropic spherical geometry
with different forms of the matter distribution and checked the role
of distinct physical terms on its collapsing rate in
$f(\mathcal{R},\mathcal{T},\mathcal{Q})$ framework. Sharif \emph{et
al.} \cite{27af,27ag} studied this concept corresponding to perfect
as well as anisotropic charged/uncharged matter sources in the
context of several modified theories. They found charge as a
significant aspect to slow down the collapse which ultimately makes
the system more stable.

This paper is based on the discussion of dynamical dissipative
cylinder that involves principal stresses in three different
directions under the influence of an electromagnetic field in
$f(\mathcal{R},\mathcal{T},\mathcal{R}_{\phi\psi}\mathcal{T}^{\phi\psi})$
scenario. The paper is organized in the following form. We present
some basic definitions of this extended theory and formulate the
field equations corresponding to the model
$\mathcal{R}+\Phi\sqrt{\mathcal{T}}+\Psi\mathcal{Q}$ in the next
section. Section \textbf{3} develops the non-vanishing dynamical
equations which are then coupled with the acceleration of the fluid
distribution. The modified transport equation as well as the
dynamical forces are constructed in section \textbf{4} to check how
these quantities affect the collapsing rate in the presence of
charge. Section \textbf{5} relates the Weyl scalar with the
effective physical quantities, the bulk viscosity and charge.
Section \textbf{6} sums up all our findings.

\section{$f(\mathcal{R},\mathcal{T},\mathcal{R}_{\varphi\vartheta}\mathcal{T}^{\varphi\vartheta})$ Theory}

The Einstein-Hilbert action takes the form after the inclusion of
modified functional $f(\mathcal{R},\mathcal{T},\mathcal{Q})$ (with
$\kappa=8\pi$) as \cite{23}
\begin{equation}\label{g1}
\mathbb{S}_{f(\mathcal{R},\mathcal{T},\mathcal{Q})}=\int\sqrt{-g}
\left\{\frac{f(\mathcal{R},\mathcal{T},\mathcal{Q})}{16\pi}+\mathbb{L}_{\mathcal{E}}
+\mathbb{L}_{\mathfrak{M}}\right\}d^{4}x,
\end{equation}
where the Lagrangian densities $\mathbb{L}_{\mathcal{E}}$ and
$\mathbb{L}_{\mathfrak{M}}$ correspond to the electromagnetic field
and the matter distribution, respectively. After implementation of
the variational principle, the action \eqref{g1} yields the field
equations as
\begin{equation}\label{g2}
\mathcal{G}_{\varphi\vartheta}=\mathcal{T}_{\varphi\vartheta}^{(\mathrm{EFF})}=\frac{1}
{f_{\mathcal{R}}-\mathbb{L}_{\mathfrak{M}}f_{\mathcal{Q}}}\left\{8\pi\big(\mathcal{T}_{\varphi\vartheta}+\mathcal{E}_{\varphi\vartheta}\big)
+\mathcal{T}_{\varphi\vartheta}^{(D)}\right\},
\end{equation}
where
$\mathcal{G}_{\varphi\vartheta},~\mathcal{T}_{\varphi\vartheta}$ and
$\mathcal{T}_{\varphi\vartheta}^{(\mathrm{EFF})}$ are labeled as the
Einstein tensor, the anisotropic matter as well as the effective
$\mathbb{EMT}$, respectively. Also, $\mathcal{E}_{\varphi\vartheta}$
is the electromagnetic field tensor. The factor
$\mathcal{T}_{\varphi\vartheta}^{(D)}$ due to the modification of
gravity has the form
\begin{eqnarray}\nonumber
\mathcal{T}_{\varphi\vartheta}^{(D)}&=&\left(f_{\mathcal{T}}+\frac{1}{2}\mathcal{R}f_{\mathcal{Q}}\right)\mathcal{T}_{\varphi\vartheta}
+\left\{\frac{\mathcal{R}}{2}\left(\frac{f}{\mathcal{R}}-f_{\mathcal{R}}\right)
-\frac{1}{2}\nabla_{\varrho}\nabla_{\omega}\big(f_{\mathcal{Q}}\mathcal{T}^{\varrho\omega}\big)\right.\\\nonumber
&-&\left.\mathbb{L}_{\mathfrak{M}}f_{\mathcal{T}}\right\}g_{\varphi\vartheta}
-\frac{1}{2}\Box\big(f_{\mathcal{Q}}\mathcal{T}_{\varphi\vartheta}\big)-2f_{\mathcal{Q}}\mathcal{R}_{\varrho(\varphi}
\mathcal{T}_{\vartheta)}^{\varrho}+\nabla_{\varrho}\nabla_{(\varphi}[\mathcal{T}_{\vartheta)}^{\varrho}
f_{\mathcal{Q}}]\\\label{g4}
&-&\big(g_{\varphi\vartheta}\Box-\nabla_{\varphi}\nabla_{\vartheta}\big)f_{\mathcal{R}}+2\big(f_{\mathcal{Q}}\mathcal{R}^{\varrho\omega}
+f_{\mathcal{T}}g^{\varrho\omega}\big)\frac{\partial^2\mathbb{L}_{\mathfrak{M}}}{\partial
g^{\varphi\vartheta}\partial g^{\varrho\omega}},
\end{eqnarray}
where $f_{\mathcal{R}}=\frac{\partial
f(\mathcal{R},\mathcal{T},\mathcal{Q})}{\partial
\mathcal{R}}$,~$f_{\mathcal{T}}=\frac{\partial
f(\mathcal{R},\mathcal{T},\mathcal{Q})}{\partial \mathcal{T}}$ and
$f_{\mathcal{Q}}=\frac{\partial
f(\mathcal{R},\mathcal{T},\mathcal{Q})}{\partial \mathcal{Q}}$.
Also, $\Box\equiv
\frac{1}{\sqrt{-g}}\partial_\varphi\big(\sqrt{-g}g^{\varphi\vartheta}\partial_{\vartheta}\big)$
and $\nabla_\varrho$ are the D'Alembert operator and the covariant
derivative, respectively. Generally, the matter Lagrangian can be in
terms of the energy density or pressure of the fluid. However, in
this case, we take its most suitable choice due to the presence of
charge as
$\mathbb{L}_{\mathfrak{M}}=-\frac{1}{4}\mathcal{H}_{\varphi\vartheta}\mathcal{H}^{\varphi\vartheta}$
which leads to $\frac{\partial^2\mathbb{L}_{\mathfrak{M}}} {\partial
g^{\varphi\vartheta}\partial
g^{\varrho\omega}}=-\frac{1}{2}\mathcal{H}_{\varphi\varrho}\mathcal{H}_{\vartheta\omega}$
\cite{22}. Here, the Maxwell field tensor
$\mathcal{H}_{\varphi\vartheta}=\varpi_{\vartheta;\varphi}-\varpi_{\varphi;\vartheta}$
is defined in terms of the four potential
$\varpi_{\vartheta}=\varpi(r)\delta_{\vartheta}^{0}$.

We consider the interior geometry representing dynamical cylindrical
structure to discuss the collapse as
\begin{equation}\label{g6}
ds^2=-\mathcal{A}^2dt^2+\mathcal{B}^2dr^2+\mathcal{C}^2d\theta^2+dz^2,
\end{equation}
where the metric potentials $\mathcal{A},~\mathcal{B}$ and
$\mathcal{C}$ are functions of $r$ and $t$. The temporal and radial
coefficients are dimensionless, while $\mathcal{C}$ has the
dimension of length. We study the considered geometry which is
coupled with anisotropic fluid in the presence of heat dissipation,
thus the most general $\mathbb{EMT}$ is given as
\begin{align}\nonumber
\mathcal{T}_{\varphi\vartheta}&=\big(\mu+\mathrm{P}_{\mathrm{r}}\big)\mathcal{U}_{\varphi}\mathcal{U}_{\vartheta}
+\mathrm{P}_{\mathrm{r}}g_{\varphi\vartheta}+\big(\mathrm{P}_{\phi}-\mathrm{P}_{\mathrm{r}}\big)
\mathcal{K}_{\varphi}\mathcal{K}_{\vartheta}+\big(\mathrm{P}_{\mathrm{z}}-\mathrm{P}_{\mathrm{r}}\big)
\mathcal{S}_{\varphi}\mathcal{S}_{\vartheta}\\\label{g5}
&+\varsigma_{\varphi}\mathcal{U}_\vartheta
+\varsigma_{\vartheta}\mathcal{U}_\varphi-\big(g_{\varphi\vartheta}+\mathcal{U}_{\varphi}\mathcal{U}_{\vartheta}\big)\zeta\Theta,
\end{align}
where $\mu$ is the energy density, $\zeta$ and $\Theta$ are the
coefficient of bulk viscosity and the expansion scalar,
respectively. The three principal stresses are indicated by
$\mathrm{P}_{\mathrm{r}},~\mathrm{P}_{\phi}$ and
$\mathrm{P}_{\mathrm{z}}$. The remaining quantities such as the
four-velocity ($\mathcal{U}_{\varphi}$), four-vectors
($\mathcal{K}_{\varphi}$ and $\mathcal{S}_{\varphi}$), $\Theta$ and
the heat flux $\varsigma_{\varphi}$ are expressed as
\begin{align}
\mathcal{U}_{\varphi}=-\mathcal{A}\delta_{\varphi}^{0},\quad
\mathcal{K}_{\varphi}=\mathcal{C}\delta_{\varphi}^{2},\quad
\mathcal{S}_{\varphi}=\delta_{\varphi}^{3},\quad
\Theta=\mathcal{U}^{\varphi}_{;\varphi},\quad
\varsigma_{\varphi}=\varsigma\mathcal{B}\delta_{\varphi}^{1},
\end{align}
obeying the following relations
\begin{align}
\mathcal{U}_{\varphi}\mathcal{U}^{\varphi}=-1,\quad
\mathcal{K}_{\varphi}\mathcal{K}^{\varphi}=1,\quad
\mathcal{S}_{\varphi} \mathcal{S}^{\varphi}=1,\quad
\mathcal{U}_{\varphi}
\mathcal{K}^{\varphi}=0=\mathcal{S}_{\varphi}\mathcal{K}^{\varphi}=\mathcal{U}_{\varphi}
\mathcal{W}^{\varphi}.
\end{align}
The coupling between matter components and geometry in any extended
gravitational theory results in the non-zero divergence of the
corresponding $\mathbb{EMT}$, i.e., $\nabla_\varphi
\mathcal{T}^{\varphi\vartheta}\neq 0$. This leads to an extra force
due to which particles start to move in non-geodesic path in their
gravitational field. Consequently, we obtain
\begin{align}\nonumber
\nabla^\varphi\big(\mathcal{T}_{\varphi\vartheta}+\mathcal{E}_{\varphi\vartheta}\big)&=\frac{2}{2f_\mathcal{T}+\mathcal{R}f_\mathcal{Q}
+16\pi}\bigg[\nabla_\varphi\big(f_\mathcal{Q}\mathcal{R}^{\varrho\varphi}\mathcal{T}_{\varrho\vartheta}\big)
+\nabla_\vartheta\big(\mathbb{L}_\mathfrak{M}f_\mathcal{T}\big)\\\nonumber
&-\mathcal{G}_{\varphi\vartheta}\nabla^\varphi\big(f_\mathcal{Q}\mathbb{L}_\mathfrak{M}\big)-\frac{1}{2}\nabla_\vartheta
\mathcal{T}^{\varrho\omega}\big(f_\mathcal{T}g_{\varrho\omega}+f_\mathcal{Q}\mathcal{R}_{\varrho\omega}\big)\\\label{g4a}
&-\frac{1}{2}\big\{\nabla^{\varphi}(\mathcal{R}f_{\mathcal{Q}})+2\nabla^{\varphi}f_{\mathcal{T}}\big\}\mathcal{T}_{\varphi\vartheta}\bigg].
\end{align}
The trace of $f(\mathcal{R},\mathcal{T},\mathcal{Q})$ field
equations is given as
\begin{align}\nonumber
&3\nabla^{\varrho}\nabla_{\varrho}
f_\mathcal{R}-\mathcal{R}\left(\frac{\mathcal{T}}{2}f_\mathcal{Q}-f_\mathcal{R}\right)-\mathcal{T}(8\pi+f_\mathcal{T})+\frac{1}{2}
\nabla^{\varrho}\nabla_{\varrho}(f_\mathcal{Q}\mathcal{T})\\\nonumber
&+\nabla_\varphi\nabla_\varrho(f_\mathcal{Q}\mathcal{T}^{\varphi\varrho})-2f+(\mathcal{R}f_\mathcal{Q}+4f_\mathcal{T})\mathbb{L}_\mathfrak{M}
+2\mathcal{R}_{\varphi\varrho}\mathcal{T}^{\varphi\varrho}f_\mathcal{Q}\\\nonumber
&-2g^{\vartheta\xi}
\frac{\partial^2\mathbb{L}_\mathfrak{M}}{\partial
g^{\vartheta\xi}\partial
g^{\varphi\varrho}}\left(f_\mathcal{T}g^{\varphi\varrho}+f_\mathcal{Q}\mathcal{R}^{\varphi\varrho}\right)=0.
\end{align}
The gravitational effects of $f(\mathcal{R},\mathcal{T})$ framework
can be achieved by disappearing $f_\mathcal{Q}$, whereas,
considering the vacuum case helps to get $f(\mathcal{R})$ gravity.

Moreover, the electromagnetic $\mathbb{EMT}$ has the form
\begin{equation*}
\mathcal{E}_{\varphi\vartheta}=\frac{1}{4\pi}\left[\frac{1}{4}g_{\varphi\vartheta}\mathcal{H}^{\varrho\xi}\mathcal{H}_{\varrho\xi}
-\mathcal{H}^{\xi}_{\varphi}\mathcal{H}_{\xi\vartheta}\right].
\end{equation*}
The Maxwell equations are
\begin{equation*}
\mathcal{H}^{\varphi\vartheta}_{;\vartheta}=4\pi
\mathcal{J}^{\varphi}, \quad
\mathcal{H}_{[\varphi\vartheta;\varrho]}=0,
\end{equation*}
which (equation on the left side) becomes for the geometry
\eqref{g6} as
\begin{equation*}
\varpi''-\bigg(\frac{\mathcal{A}'}{\mathcal{A}}+\frac{\mathcal{B}'}{\mathcal{B}}
-\frac{\mathcal{C}'}{\mathcal{C}}\bigg)\varpi'=4\pi\rho\mathcal{AB}^2,
\end{equation*}
yielding
\begin{equation}\nonumber
\varpi'=\frac{s\mathcal{AB}}{\mathcal{C}}.
\end{equation}
Here, $\mathcal{J}^{\varphi}=\rho \mathcal{K}^{\varphi}$,
$\mathcal{J}^{\varphi}$ and $\rho$ are the current and charge
densities, respectively. Also, the presence of charge in the
interior distribution is indicated by $s=4\pi\int\rho\mathcal{BC}dr$
and $'=\frac{\partial}{\partial r}$. Consequently, the matter
Lagrangian becomes
$\mathbb{L}_{\mathfrak{M}}=\frac{s^2}{2\mathcal{C}^2}$.

We consider a linear model to study the current setup as
\begin{equation}\label{g5d}
f(\mathcal{R},\mathcal{T},\mathcal{Q})=f_1(\mathcal{R})+f_2(\mathcal{T})+f_3(\mathcal{Q})=\mathcal{R}+\Phi\sqrt{\mathcal{T}}+\Psi\mathcal{Q}.
\end{equation}
It is noteworthy that the values of the coupling parameters
involving in any gravitational model should be within their observed
ranges to get physically acceptable results. For $\Phi=0$, the above
model has been studied through which some stable isotropic
configurations are obtained by choosing different values of $\Psi$
\cite{22a,22b}. The terms $\mathcal{R},~\mathcal{T}$ and
$\mathcal{Q}$ in this case become
\begin{align}\nonumber
\mathcal{R}&=-\frac{2}{\mathcal{A}^3\mathcal{B}^3\mathcal{C}}\bigg[\mathcal{A}^3\mathcal{B}\mathcal{C}''
-\mathcal{A}\mathcal{B}^3\ddot{\mathcal{C}}-\mathcal{A}\mathcal{B}^2\mathcal{C}\ddot{\mathcal{B}}+\mathcal{A}^2\mathcal{B}\mathcal{C}\mathcal{A}''
-\mathcal{A}^3\mathcal{B}'\mathcal{C}'+\mathcal{B}^3\dot{\mathcal{A}}\dot{\mathcal{C}}\\\nonumber
&+\mathcal{A}^2\mathcal{B}\mathcal{A}'\mathcal{C}'-\mathcal{A}\mathcal{B}^2\dot{\mathcal{B}}\dot{\mathcal{C}}
+\mathcal{B}^2\mathcal{C}\dot{\mathcal{A}}\dot{\mathcal{B}}-\mathcal{A}^2\mathcal{C}\mathcal{A}'\mathcal{B}'\bigg],\\\nonumber
\mathcal{T}&=-\mu+\mathrm{P}_{\mathrm{r}}+\mathrm{P}_{\phi}+\mathrm{P}_{\mathrm{z}}-3\zeta\Theta,\\\nonumber
\mathcal{Q}&=-\frac{1}{\mathcal{A}^3\mathcal{B}^3\mathcal{C}}\bigg[\mu\big\{\mathcal{A}\mathcal{B}^2\mathcal{C}\ddot{\mathcal{B}}
-\mathcal{A}^2\mathcal{B}\mathcal{C}\mathcal{A}''+\mathcal{A}^2\mathcal{CA}'\mathcal{B}'-\mathcal{A}^2\mathcal{BA}'\mathcal{C}'
+\mathcal{AB}^3\ddot{\mathcal{C}}\\\nonumber
&-\mathcal{B}^2\mathcal{C}\dot{\mathcal{A}}\dot{\mathcal{B}}-\mathcal{B}^3\dot{\mathcal{A}}\dot{\mathcal{C}}\big\}+2\varsigma
\mathcal{AB}\big\{\mathcal{AB}\dot{\mathcal{C}}'-\mathcal{BA}'\dot{\mathcal{C}}-\mathcal{A}\dot{\mathcal{B}}\mathcal{C}'\big\}
+\big(\mathrm{P}_{\mathrm{r}}-\zeta\Theta\big)\\\nonumber
&\big\{\mathcal{B}^2\mathcal{C}\dot{\mathcal{A}}\dot{\mathcal{B}}-\mathcal{AB}^2\mathcal{C}\ddot{\mathcal{B}}+\mathcal{A}^2\mathcal{BCA}''
-\mathcal{AB}^2\dot{\mathcal{B}}\dot{\mathcal{C}}-\mathcal{A}^3\mathcal{B}'\mathcal{C}'-\mathcal{A}^2\mathcal{CA}'\mathcal{B}'\\\nonumber
&+\mathcal{A}^3\mathcal{BC}''\big\}+\big(\mathrm{P}_{\phi}-\zeta\Theta\big)\big\{\mathcal{A}^3\mathcal{BC}''
-\mathcal{A}^3\mathcal{B}'\mathcal{C}'+\mathcal{A}^2\mathcal{BA}'\mathcal{C}'-\mathcal{AB}^2\dot{\mathcal{B}}\dot{\mathcal{C}}\\\nonumber
&-\mathcal{AB}^3\ddot{\mathcal{C}}+\mathcal{B}^3\dot{\mathcal{A}}\dot{\mathcal{C}}\big\}\bigg],
\end{align}
where $.=\frac{\partial}{\partial t}$. The field equations
corresponding to the cylindrical spacetime are
\begin{align}\label{g8}
\frac{1}{1-\frac{\Psi{s^2}}{2\mathcal{C}^2}}\left(8\pi\mu+\frac{s^2}{\mathcal{C}^2}+\frac{\mu^{(D)}}{\mathcal{A}^{2}}
+\frac{\mathcal{E}_{00}^{(D)}}{\mathcal{A}^{2}}\right)&=\frac{\mathcal{B}'\mathcal{C}'}{\mathcal{B}^{3}\mathcal{C}}
-\frac{\mathcal{C}''}{\mathcal{B}^{2}\mathcal{C}}+\frac{\dot{\mathcal{B}}\dot{\mathcal{C}}}{\mathcal{A}^{2}\mathcal{BC}},\\\label{g8a}
\frac{1}{1-\frac{\Psi{s^2}}{2\mathcal{C}^2}}\left(8\pi\mathrm{P}_{\mathrm{r}}-\frac{s^2}{\mathcal{C}^2}-\zeta\Theta
+\frac{\mathrm{P}_{\mathrm{r}}^{(D)}}{\mathcal{B}^{2}}+\frac{\mathcal{E}_{11}^{(D)}}{\mathcal{B}^{2}}\right)&=
\frac{\dot{\mathcal{A}}\dot{\mathcal{C}}}{\mathcal{A}^{3}\mathcal{C}}-\frac{\ddot{\mathcal{C}}}{\mathcal{A}^{2}\mathcal{C}}
+\frac{\mathcal{A}'\mathcal{C}'}{\mathcal{A}\mathcal{B}^{2}\mathcal{C}},\\\label{g8b}
\frac{1}{1-\frac{\Psi{s^2}}{2\mathcal{C}^2}}\left(8\pi\mathrm{P}_{\phi}+\frac{s^2}{\mathcal{C}^2}-\zeta\Theta
+\frac{\mathrm{P}_{\phi}^{(D)}}{\mathcal{C}^{2}}\right)&=
\frac{\dot{\mathcal{A}}\dot{\mathcal{B}}}{\mathcal{A}^{3}\mathcal{B}}+\frac{\mathcal{A}''}{\mathcal{A}\mathcal{B}^{2}}
-\frac{\ddot{\mathcal{B}}}{\mathcal{A}^{2}\mathcal{B}}-\frac{\mathcal{A}'\mathcal{B}'}{\mathcal{A}\mathcal{B}^{3}},\\\nonumber
\frac{1}{1-\frac{\Psi{s^2}}{2\mathcal{C}^2}}\left(8\pi\mathrm{P}_{\mathrm{z}}+\frac{s^2}{\mathcal{C}^2}-\zeta\Theta
+\mathrm{P}_{\mathrm{z}}^{(D)}\right)&=\frac{\dot{\mathcal{A}}\dot{\mathcal{C}}}{\mathcal{A}^{3}\mathcal{C}}
-\frac{\ddot{\mathcal{B}}}{\mathcal{A}^{2}\mathcal{B}}-\frac{\ddot{\mathcal{C}}}{A^{2}\mathcal{C}}
+\frac{\dot{\mathcal{A}}\dot{\mathcal{B}}}{\mathcal{A}^{3}\mathcal{B}}\\\nonumber
&+\frac{\mathcal{A}''}{\mathcal{A}\mathcal{B}^{2}}+\frac{\mathcal{C}''}{\mathcal{B}^{2}\mathcal{C}}
+\frac{\mathcal{A}'\mathcal{C}'}{\mathcal{A}\mathcal{B}^{2}\mathcal{C}}-\frac{\mathcal{A}'\mathcal{B}'}{\mathcal{A}\mathcal{B}^{3}}\\\label{g8c}
&-\frac{\mathcal{B}'\mathcal{C}'}{\mathcal{B}^{3}\mathcal{C}}-\frac{\dot{\mathcal{B}}\dot{\mathcal{C}}}{\mathcal{A}^{2}\mathcal{B}\mathcal{C}},
\\\label{g8d}\frac{1}{1-\frac{\Psi{s^2}}{2\mathcal{C}^2}}\left(8\pi\varsigma-\frac{\varsigma^{(D)}}{\mathcal{AB}}
-\frac{\mathcal{E}_{01}^{(D)}}{\mathcal{AB}}\right)&=\frac{\dot{\mathcal{C}}'}{\mathcal{A}\mathcal{B}\mathcal{C}}
-\frac{\dot{\mathcal{B}}\mathcal{C}'}{\mathcal{A}\mathcal{B}^{2}\mathcal{C}}
-\frac{\mathcal{A}'\dot{\mathcal{C}}}{\mathcal{A}^{2}\mathcal{B}\mathcal{C}},
\end{align}
which describe how matter is coupled with gravity. Due to the
modification of gravity, the quantities
$\mu^{(D)},~\mathrm{P}_{\mathrm{r}}^{(D)},~\mathrm{P}_{\phi}^{(D)},~\mathrm{P}_{\mathrm{z}}^{(D)},~\varsigma^{(D)},
~\mathcal{E}_{00}^{(D)},~\mathcal{E}_{11}^{(D)}$ and
$\mathcal{E}_{01}^{(D)}$ appear on the left hand side of the above
equations and we provide their values in Appendix \textbf{A}. The
quantities
$\left(8\pi\mu+\frac{s^2}{\mathcal{C}^2}+\frac{\mu^{(D)}}{\mathcal{A}^{2}}+\frac{\mathcal{E}_{00}^{(D)}}{\mathcal{A}^{2}}\right)$,
$\left(8\pi\varsigma-\frac{\varsigma^{(D)}}{\mathcal{AB}}-\frac{\mathcal{E}_{01}^{(D)}}{\mathcal{AB}}\right)$,
$\left(8\pi\mathrm{P}_{\mathrm{r}}-\frac{s^2}{\mathcal{C}^2}-\zeta\Theta
+\frac{\mathrm{P}_{\mathrm{r}}^{(D)}}{\mathcal{B}^{2}}+\frac{\mathcal{E}_{11}^{(D)}}{\mathcal{B}^{2}}\right)$,
$\left(8\pi\mathrm{P}_{\phi}+\frac{s^2}{\mathcal{C}^2}-\zeta\Theta
+\frac{\mathrm{P}_{\phi}^{(D)}}{\mathcal{C}^{2}}\right)$ and
$\left(8\pi\mathrm{P}_{\mathrm{z}}+\frac{s^2}{\mathcal{C}^2}-\zeta\Theta+\mathrm{P}_{\mathrm{z}}^{(D)}\right)$
depict the effective energy density, effective heat flux and the
effective principal pressures, respectively.

Thorne \cite{41ba} provides the formula of C-energy to determine the
inner mass distribution as
\begin{equation}\label{g13}
\tilde{m}(t,r)=\mathfrak{L}\hat{\mathrm{E}}=\frac{\mathfrak{L}}{8}(1-\mathfrak{L}^{-2}\nabla_\varphi
\hat{r}\nabla^\varphi \hat{r}),
\end{equation}
where $\hat r=\varrho\mathfrak{L}$ symbolizes the circumference
radius, $\varrho$ is the areal radius and $\mathfrak{L}$ is the
specific length which are defined as
\begin{equation}\nonumber
\varrho^2=\eta_{(1)\vartheta}\eta_{(1)}^{\vartheta}, \quad
\mathfrak{L}^2=\eta_{(2)\vartheta}\eta_{(2)}^{\vartheta},
\end{equation}
along with the Killing vectors $\eta_{(1)}=\frac{\partial}{\partial
\theta}$ and $\eta_{(2)}=\frac{\partial}{\partial z}$. Also, $\hat
{\mathrm{E}}$ is the gravitational energy per specific length. After
some manipulation, Eq.\eqref{g13} yields the mass as
\begin{equation}\label{g14}
\tilde{m}=\frac{\mathfrak{L}}{8}\bigg[1-\bigg(\frac{C'}{B}\bigg)^2+\bigg(\frac{\dot
C}{A}\bigg)^2+\frac{s^2}{2}\bigg].
\end{equation}

\section{Dynamics of the Cylindrical Star}

Initially, the evolution of spherical geometry was studied by Misner
and Sharp through the formulation of certain dynamical quantities.
The velocity and acceleration of the collapsing source (under
consideration) was calculated by using proper radial as well as
temporal derivatives. Later, these definitions have extensively been
employed to explore the evolutionary pattern of spherical and
cylindrical spacetimes \cite{41bb}. In this modified scenario, these
equations are
\begin{align}\label{g21}
\mathcal{T}_{\varphi;\vartheta}^{(\mathrm{EFF})\vartheta}\mathcal{U}^{\varphi}&=\big(8\pi\mathcal{T}_{\varphi}^{\vartheta}
+\mathcal{T}_{\varphi}^{(D)\vartheta}\big)_{;\vartheta}\mathcal{U}^{\varphi}=0,\\\label{g22}
\mathcal{T}_{\varphi;\vartheta}^{(\mathrm{EFF})\vartheta}\varsigma^{\varphi}&=\big(8\pi\mathcal{T}_{\varphi}^{\vartheta}
+\mathcal{T}_{\varphi}^{(D)\vartheta}\big)_{;\vartheta}\varsigma^{\varphi}=0.
\end{align}
The above equations yield, respectively, as
\begin{align}\nonumber
&\frac{1}{\mathcal{A}^2}\bigg(8\pi\mu+\frac{s^2}{\mathcal{C}^2}+\frac{\mu^{(D)}}{\mathcal{A}^{2}}
+\frac{\mathcal{E}_{00}^{(D)}}{\mathcal{A}^{2}}\bigg)^.+\frac{\dot{\mathcal{B}}}{\mathcal{A}^2\mathcal{B}}
\bigg(8\pi\mu+\frac{\mu^{(D)}}{\mathcal{A}^{2}}+\frac{\mathcal{E}_{00}^{(D)}}{\mathcal{A}^{2}}+8\pi\mathrm{P}_{\mathrm{r}}\\\nonumber
&-\zeta\Theta+\frac{\mathrm{P}_{\mathrm{r}}^{(D)}}{\mathcal{B}^{2}}+\frac{\mathcal{E}_{11}^{(D)}}{\mathcal{B}^{2}}\bigg)
+\frac{1}{\mathcal{AB}}\bigg(8\pi\varsigma-\frac{\varsigma^{(D)}}{\mathcal{AB}}-\frac{\mathcal{E}_{01}^{(D)}}{\mathcal{AB}}\bigg)'
+\frac{\dot{\mathcal{C}}}{\mathcal{A}^2\mathcal{C}}\bigg(8\pi\mu+\frac{2s^2}{\mathcal{C}^2}\\\nonumber
&+\frac{\mu^{(D)}}{\mathcal{A}^{2}}+\frac{\mathcal{E}_{00}^{(D)}}{\mathcal{A}^{2}}+8\pi\mathrm{P}_{\phi}-\zeta\Theta
+\frac{\mathrm{P}_{\phi}^{(D)}}{\mathcal{C}^{2}}\bigg)+\frac{1}{\mathcal{AB}}
\bigg(8\pi\varsigma-\frac{\varsigma^{(D)}}{\mathcal{AB}}-\frac{\mathcal{E}_{01}^{(D)}}{\mathcal{AB}}\bigg)\\\label{g23}
&\times\bigg(\frac{2\mathcal{A}'}{\mathcal{A}}+\frac{\mathcal{C}'}{\mathcal{C}}\bigg)=0,\\\nonumber
&\frac{\mathcal{B}}{\mathcal{A}}\bigg(8\pi\varsigma-\frac{\varsigma^{(D)}}{\mathcal{AB}}-\frac{\mathcal{E}_{01}^{(D)}}{\mathcal{AB}}\bigg)^.
+\bigg(8\pi\mathrm{P}_{\mathrm{r}}-\frac{s^2}{\mathcal{C}^2}-\zeta\Theta+\frac{\mathrm{P}_{\mathrm{r}}^{(D)}}{\mathcal{B}^{2}}
+\frac{\mathcal{E}_{11}^{(D)}}{\mathcal{B}^{2}}\bigg)'\\\nonumber
&+\frac{\mathcal{A}'}{\mathcal{A}}\bigg(8\pi\mu+\frac{\mu^{(D)}}{\mathcal{A}^{2}}+\frac{\mathcal{E}_{00}^{(D)}}{\mathcal{A}^{2}}
+8\pi\mathrm{P}_{\mathrm{r}}-\zeta\Theta+\frac{\mathrm{P}_{\mathrm{r}}^{(D)}}{\mathcal{B}^{2}}
+\frac{\mathcal{E}_{11}^{(D)}}{\mathcal{B}^{2}}\bigg)+\frac{\mathcal{C}'}{\mathcal{C}}\\\nonumber
&\times\bigg(8\pi\mathrm{P}_{\mathrm{r}}-\frac{2s^2}{\mathcal{C}^2}+\frac{\mathrm{P}_{\mathrm{r}}^{(D)}}{\mathcal{B}^{2}}
+\frac{\mathcal{E}_{11}^{(D)}}{\mathcal{B}^{2}}-8\pi\mathrm{P}_{\phi}-\frac{\mathrm{P}_{\phi}^{(D)}}{\mathcal{C}^{2}}\bigg)
+\frac{\mathcal{B}}{\mathcal{A}}\bigg(\frac{2\dot{\mathcal{B}}}{\mathcal{B}}+\frac{\dot{\mathcal{C}}}{\mathcal{C}}\bigg)\\\label{g24}
&\times\bigg(8\pi\varsigma-\frac{\varsigma^{(D)}}{\mathcal{AB}}-\frac{\mathcal{E}_{01}^{(D)}}{\mathcal{AB}}\bigg)=0.
\end{align}
Different variations occurring  in the structural evolution of
self-gravitating objects can be studied with the help of these
significant equations. Some important terms play a vital role in the
discussion of dynamics of the collapsing source. We define proper
radial as well as temporal derivatives as \cite{29a,41bb}
\begin{equation}\label{g25}
\mathfrak{D}_{\mathrm{r}}=\frac{1}{\mathcal{C}^{\prime}}
\frac{\partial}{\partial \mathrm{r}}, \quad
\mathfrak{D}_{\mathrm{t}}=\frac{1}{\mathcal{A}}
\frac{\partial}{\partial \mathrm{t}}.
\end{equation}

The reduction of outward-directed pressure of a compact star allows
gravity to dominate which results in the collapse of that object.
Consequently, the radius continuously decreases and the velocity of
interior distribution becomes negative, i.e.,
\begin{equation}\label{g26}
\mathbb{U}=\mathfrak{D}_{\mathrm{t}}(\mathcal{C})=\frac{\dot{\mathcal{C}}}{\mathcal{A}}<0.
\end{equation}
Combining this equation with C-energy \eqref{g14}, we obtain
\begin{equation}\label{g27}
\frac{\mathcal{C}'}{\mathcal{B}}=\left(1+\mathbb{U}^{2}-\frac{8
\tilde{m}}{\mathfrak{L}}+\frac{s^2}{2}\right)^{\frac{1}{2}}=\omega.
\end{equation}

Applying the definition of temporal derivative
($\mathfrak{D}_{\mathrm{t}}$) on the C-energy, we have the following
form
\begin{align}\nonumber
\mathfrak{D}_{\mathrm{t}}(\tilde{m})&=-\frac{\mathcal{C}\mathfrak{L}}{4(1-\frac{\Psi{s^2}}{2\mathcal{C}^2})}
\left\{\left(8\pi\mathrm{P}_{\mathrm{r}}-\frac{s^2}{\mathcal{C}^2}-\zeta\Theta
+\frac{\mathrm{P}_{\mathrm{r}}^{(D)}}{\mathcal{B}^{2}}+\frac{\mathcal{E}_{11}^{(D)}}{\mathcal{B}^{2}}\right)\mathbb{U}\right.\\\label{g28}
&\left.+\left(8\pi\varsigma-\frac{\varsigma^{(D)}}{\mathcal{AB}}-\frac{\mathcal{E}_{01}^{(D)}}{\mathcal{AB}}\right)\omega\right\},
\end{align}
which explains the variation of total energy with respect to time.
Equation \eqref{g28} further demonstrates that how the expansion
scalar and effective forms of radial pressure as well as heat flux
and the electromagnetic field influence the collapsing phenomenon.
The increment in the total energy is guaranteed from the factor
$\left(8\pi\mathrm{P}_{\mathrm{r}}-\frac{s^2}{\mathcal{C}^2}-\zeta\Theta
+\frac{\mathrm{P}_{\mathrm{r}}^{(D)}}{\mathcal{B}^{2}}+\frac{\mathcal{E}_{11}^{(D)}}{\mathcal{B}^{2}}\right)\mathbb{U}$
(appears on the right hand side of above equation), that becomes
positive due to the negative velocity $\mathbb{U}$. On the other
hand, the dissipation of heat from the considered source can be
observed through the entity
$\left(8\pi\varsigma-\frac{\varsigma^{(D)}}{\mathcal{AB}}-\frac{\mathcal{E}_{01}^{(D)}}{\mathcal{AB}}\right)\omega$,
which consequently reduces the total energy.

Next, we take the proper radial derivative of Eq.\eqref{g14} and
couple it with Eqs.\eqref{g8} and \eqref{g8d} to discuss that how
energy between the adjoining cylindrical surfaces varies, as
\begin{align}\nonumber
\mathfrak{D}_{\mathrm{r}}(\tilde{m})&=\frac{\mathfrak{L}}{4}\left[\frac{\mathcal{C}}{1-\frac{\Psi{s^2}}{2\mathcal{C}^2}}\left\{\left(8\pi\mu
+\frac{s^2}{\mathcal{C}^2}+\frac{\mu^{(D)}}{\mathcal{A}^{2}}+\frac{\mathcal{E}_{00}^{(D)}}{\mathcal{A}^{2}}\right)\right.\right.\\\label{g29}
&+\left.\left.\left(8\pi\varsigma-\frac{\varsigma^{(D)}}{\mathcal{AB}}-\frac{\mathcal{E}_{01}^{(D)}}{\mathcal{AB}}\right)
\frac{\mathbb{U}}{\omega}\right\}+\frac{ss'}{2\mathcal{C}^{'}}\right].
\end{align}
The quantity
$\left(8\pi\mu+\frac{s^2}{\mathcal{C}^2}+\frac{\mu^{(D)}}{\mathcal{A}^{2}}+\frac{\mathcal{E}_{00}^{(D)}}{\mathcal{A}^{2}}\right)$
on the right side indicates that the effective energy density
affects the collapse rate of the considered matter source and
ultimately increases its total energy. The dissipation of heat from
the geometry is guaranteed by the second entity
$\left(8\pi\varsigma-\frac{\varsigma^{(D)}}{\mathcal{AB}}-\frac{\mathcal{E}_{01}^{(D)}}{\mathcal{AB}}\right)\frac{\mathbb{U}}{\omega}$
due to negative velocity of the fluid. The last term exposes the
effect of Coulomb force on the system. The implementation of
$\mathfrak{D}_{\mathrm{t}}$ on velocity of the collapsing source
provides the acceleration as
\begin{align}\nonumber
\mathfrak{D}_{\mathrm{t}}(\mathbb{U})&=-\frac{\mathcal{C}}{1-\frac{\Psi{s^2}}{2\mathcal{C}^2}}\bigg(8\pi\mathrm{P}_{\mathrm{r}}
-\frac{s^2}{\mathcal{C}^2}-\zeta\Theta+\frac{\mathrm{P}_{\mathrm{r}}^{(D)}}{\mathcal{B}^{2}}
+\frac{\mathcal{E}_{11}^{(D)}}{\mathcal{B}^{2}}\bigg)\\\label{g30}
&-\frac{\tilde{m}}{\mathcal{C}^2}+\frac{\omega\mathcal{A}'}{\mathcal{AB}}+\frac{\mathfrak{L}}{8\mathcal{C}^{2}}\bigg(1+\mathbb{U}^{2}
-\omega^{2}+\frac{s^2}{2}\bigg).
\end{align}
The value of $\frac{\mathcal{A}'}{\mathcal{A}}$ can be obtained from
Eq.\eqref{g24} as
\begin{align}\nonumber
\frac{\mathcal{A}'}{\mathcal{A}}&=-\frac{1}{\left(8\pi\mu+\frac{\mu^{(D)}}{\mathcal{A}^{2}}+\frac{\mathcal{E}_{00}^{(D)}}{\mathcal{A}^{2}}
+8\pi\mathrm{P}_{\mathrm{r}}-\zeta\Theta
+\frac{\mathrm{P}_{\mathrm{r}}^{(D)}}{\mathcal{B}^{2}}+\frac{\mathcal{E}_{11}^{(D)}}{\mathcal{B}^{2}}\right)}
\bigg\{\frac{\mathcal{B}}{\mathcal{A}}\bigg(\frac{2\dot{\mathcal{B}}}{\mathcal{B}}+\frac{\dot{\mathcal{C}}}{\mathcal{C}}\bigg)\\\nonumber
&\times\bigg(8\pi\varsigma-\frac{\varsigma^{(D)}}{\mathcal{AB}}-\frac{\mathcal{E}_{01}^{(D)}}{\mathcal{AB}}\bigg)
+\bigg(8\pi\mathrm{P}_{\mathrm{r}}-\frac{s^2}{\mathcal{C}^2}-\zeta\Theta
+\frac{\mathrm{P}_{\mathrm{r}}^{(D)}}{\mathcal{B}^{2}}+\frac{\mathcal{E}_{11}^{(D)}}{\mathcal{B}^{2}}\bigg)'\\\nonumber
&+\frac{\mathcal{C}'}{\mathcal{C}}\bigg(8\pi\mathrm{P}_{\mathrm{r}}-\frac{2s^2}{\mathcal{C}^2}
+\frac{\mathrm{P}_{\mathrm{r}}^{(D)}}{\mathcal{B}^{2}}+\frac{\mathcal{E}_{11}^{(D)}}{\mathcal{B}^{2}}
-8\pi\mathrm{P}_{\phi}-\frac{\mathrm{P}_{\phi}^{(D)}}{\mathcal{C}^{2}}\bigg)+\frac{\mathcal{B}}{\mathcal{A}}\\\label{g31}
&\times\bigg(8\pi\varsigma-\frac{\varsigma^{(D)}}{\mathcal{AB}}
-\frac{\mathcal{E}_{01}^{(D)}}{\mathcal{AB}}\bigg)^.\bigg\}.
\end{align}
Inserting this in Eq.\eqref{g30}, it follows that
\begin{align}\nonumber
&\mathfrak{D}_{\mathrm{t}}(\mathbb{U})\bigg(8\pi\mu+\frac{\mu^{(D)}}{\mathcal{A}^{2}}+\frac{\mathcal{E}_{00}^{(D)}}{\mathcal{A}^{2}}
+8\pi\mathrm{P}_{\mathrm{r}}-\zeta\Theta+\frac{\mathrm{P}_{\mathrm{r}}^{(D)}}{\mathcal{B}^{2}}
+\frac{\mathcal{E}_{11}^{(D)}}{\mathcal{B}^{2}}\bigg)=-\bigg\{\frac{\tilde{m}}{\mathcal{C}^{2}}-\frac{\mathfrak{L}}{8\mathcal{C}^{2}}\\\nonumber
&\times\bigg(1+\mathbb{U}^{2}+\frac{s^2}{2}\bigg)+\frac{\mathcal{C}}{1-\frac{\Psi{s^2}}{2\mathcal{C}^2}}\bigg(8\pi\mathrm{P}_{\mathrm{r}}
-\frac{s^2}{\mathcal{C}^2}-\zeta\Theta+\frac{\mathrm{P}_{\mathrm{r}}^{(D)}}{\mathcal{B}^{2}}
+\frac{\mathcal{E}_{11}^{(D)}}{\mathcal{B}^{2}}\bigg)\bigg\}\\\nonumber
&\bigg(8\pi\mu+\frac{\mu^{(D)}}{\mathcal{A}^{2}}+\frac{\mathcal{E}_{00}^{(D)}}{\mathcal{A}^{2}}+8\pi\mathrm{P}_{\mathrm{r}}-\zeta\Theta
+\frac{\mathrm{P}_{\mathrm{r}}^{(D)}}{\mathcal{B}^{2}}+\frac{\mathcal{E}_{11}^{(D)}}{\mathcal{B}^{2}}\bigg)
-\frac{\omega^{2}}{\mathcal{C}}\bigg\{\bigg(8\pi\mathrm{P}_{\mathrm{r}}-\frac{2s^2}{\mathcal{C}^2}\\\nonumber
&+\frac{\mathrm{P}_{\mathrm{r}}^{(D)}}{\mathcal{B}^{2}}+\frac{\mathcal{E}_{11}^{(D)}}{\mathcal{B}^{2}}
-8\pi\mathrm{P}_{\phi}-\frac{\mathrm{P}_{\phi}^{(D)}}{\mathcal{C}^{2}}\bigg)
+\frac{\mathfrak{L}}{8\mathcal{C}}\bigg(8\pi\mu+\frac{\mu^{(D)}}{\mathcal{A}^{2}}
+\frac{\mathcal{E}_{00}^{(D)}}{\mathcal{A}^{2}}+8\pi\mathrm{P}_{\mathrm{r}}\\\nonumber
&-\zeta\Theta+\frac{\mathrm{P}_{\mathrm{r}}^{(D)}}{\mathcal{B}^{2}}+\frac{\mathcal{E}_{11}^{(D)}}{\mathcal{B}^{2}}\bigg)\bigg\}
-\omega\bigg\{\frac{1}{\mathcal{B}}\bigg(8\pi\mathrm{P}_{\mathrm{r}}-\frac{s^2}{\mathcal{C}^2}-\zeta\Theta
+\frac{\mathrm{P}_{\mathrm{r}}^{(D)}}{\mathcal{B}^{2}}+\frac{\mathcal{E}_{11}^{(D)}}{\mathcal{B}^{2}}\bigg)'\\\label{g32}
&+\mathfrak{D}_{\mathrm{t}}\bigg(8\pi\varsigma-\frac{\varsigma^{(D)}}{\mathcal{AB}}-\frac{\mathcal{E}_{01}^{(D)}}{\mathcal{AB}}\bigg)
+\frac{1}{\mathcal{A}}\bigg(8\pi\varsigma-\frac{\varsigma^{(D)}}{\mathcal{AB}}-\frac{\mathcal{E}_{01}^{(D)}}{\mathcal{AB}}\bigg)
\bigg(\frac{2\dot{\mathcal{B}}}{\mathcal{B}}+\frac{\dot{\mathcal{C}}}{\mathcal{C}}\bigg)\bigg\}.
\end{align}

The product of acceleration ($\mathfrak{D}_{\mathrm{t}}\mathbb{U}$)
with the inertial mass density
$\bigg(8\pi\mu+\frac{\mu^{(D)}}{\mathcal{A}^{2}}+\frac{\mathcal{E}_{00}^{(D)}}{\mathcal{A}^{2}}
+8\pi\mathrm{P}_{\mathrm{r}}-\zeta\Theta+\frac{\mathrm{P}_{\mathrm{r}}^{(D)}}{\mathcal{B}^{2}}
+\frac{\mathcal{E}_{11}^{(D)}}{\mathcal{B}^{2}}\bigg)$ on the left
side of this equation indicates the Newtonian force. At the same
time, the later term also appears in the first term on the right
side, now presenting the gravitational mass density. This ultimately
results in the fulfillment of the equivalence principle, as both the
masses are indistinguishable. The expression multiplied by
$\omega^2$ helps to determine that how the collapse rate is affected
by the gravitational mass density, electric field intensity and
effective pressures in different directions. The first term in the
last curly bracket presents the contribution of gradient of
effective radial stress and the expansion scalar in this framework.
Moreover, the hydrodynamical properties of the charged cylinder can
be well described by the last two terms involving the heat flux
along with modified corrections.

\section{Transport Equations}

In view of the $\mathbb{EMT}$ \eqref{g5} involving heat flux, some
equations play significant role to discuss the structural changes in
the interior of compact bodies, one of them is the transport
equation. Such equation discloses how one can evaluate different
physical terms such as mass, momentum and heat during the collapse.
The transport equation is given in the following which supports the
diffusion process as
\begin{equation}\label{g33}
\varrho\mathrm{h}^{\varphi\vartheta}\mathcal{U}^{\gamma}\bar{\varsigma}_{\vartheta;\gamma}+\bar{\varsigma}^{\varphi}=
-\eta\mathrm{h}^{\varphi\vartheta}\left(\tau_{,\vartheta}+\tau\mathrm{a}_{\vartheta}\right)-\frac{1}{2}\eta\tau^{2}\left(\frac{\varrho
\mathcal{U}^{\vartheta}}{\eta\tau^{2}}\right)_{;\vartheta}\bar{\varsigma}^{\varphi},
\end{equation}
where
$\bar{\varsigma}=\left(8\pi\varsigma-\frac{\varsigma^{(D)}}{\mathcal{AB}}-\frac{\mathcal{E}_{01}^{(D)}}{\mathcal{AB}}\right)$
and
$\mathrm{h}^{\varphi\vartheta}=g^{\varphi\vartheta}+\mathcal{U}^{\varphi}
\mathcal{U}^{\vartheta}$ is the projection tensor. Also,
$\varrho,~\eta,~\tau$ and $\mathrm{a}_{\vartheta}$ \big(i.e.,
$\mathrm{a}_1=\frac{\mathcal{A}'}{\mathcal{A}}$\big) are
mathematical notations of the relaxation time, thermal conductivity,
temperature and four-acceleration, respectively. Equation
\eqref{g33} yields after some simplification as
\begin{align}\nonumber
\mathcal{B}
\mathfrak{D}_{\mathrm{t}}\left(8\pi\varsigma-\frac{\varsigma^{(D)}}{\mathcal{AB}}-\frac{\mathcal{E}_{01}^{(D)}}{\mathcal{AB}}\right)&=
-\frac{\eta\tau^{2}\mathcal{B}}{2\mathcal{A}\varrho}\left(\frac{\varrho}{\eta\tau^{2}}\right)^.
\left(8\pi\varsigma-\frac{\varsigma^{(D)}}{\mathcal{AB}}-\frac{\mathcal{E}_{01}^{(D)}}{\mathcal{AB}}\right)\\\nonumber
&-\frac{\eta\tau'}{\varrho}-\frac{\eta\tau}{\varrho}\left(\frac{\mathcal{A}'}{\mathcal{A}}\right)-\frac{\mathcal{B}}{2\mathcal{A}}
\left(\frac{3\dot{\mathcal{B}}}{\mathcal{B}}+\frac{\dot{\mathcal{C}}}{\mathcal{C}}+\frac{2\mathcal{A}}{\varrho}\right)\\\label{g34}
&\times\left(8\pi\varsigma-\frac{\varsigma^{(D)}}{\mathcal{AB}}-\frac{\mathcal{E}_{01}^{(D)}}{\mathcal{AB}}\right),
\end{align}
which becomes after combining with Eq.\eqref{g31} as
\begin{align}\nonumber
\mathcal{B}
\mathfrak{D}_{\mathrm{t}}\left(8\pi\varsigma-\frac{\varsigma^{(D)}}{\mathcal{AB}}-\frac{\mathcal{E}_{01}^{(D)}}{\mathcal{AB}}\right)&=
-\frac{\eta\tau^{2}\mathcal{B}}{2\mathcal{A}\varrho}\left(\frac{\varrho}{\eta\tau^{2}}\right)^.\left(8\pi\varsigma
-\frac{\varsigma^{(D)}}{\mathcal{AB}}-\frac{\mathcal{E}_{01}^{(D)}}{\mathcal{AB}}\right)\\\nonumber
&-\frac{\mathcal{B}}{2\mathcal{A}}\left(8\pi\varsigma-\frac{\varsigma^{(D)}}{\mathcal{AB}}-\frac{\mathcal{E}_{01}^{(D)}}{\mathcal{AB}}\right)
\bigg(\frac{3\dot{\mathcal{B}}}{\mathcal{B}}+\frac{\dot{\mathcal{C}}}{\mathcal{C}}+\frac{2\mathcal{A}}{\varrho}\bigg)\\\nonumber
&-\frac{\eta\tau'}{\varrho}-\frac{\eta\tau\mathcal{B}}{\varrho\omega}\bigg\{\frac{\tilde{m}}{\mathcal{C}^2}-\frac{\mathfrak{L}}{8\mathcal{C}^{2}}
\bigg(1+\mathbb{U}^{2}-\omega^{2}+\frac{s^2}{2}\bigg)\\\label{g35}
&+\frac{\mathcal{C}}{1-\frac{\Psi{s^2}}{2\mathcal{C}^2}}\bigg(8\pi\mathrm{P}_{\mathrm{r}}-\frac{s^2}{\mathcal{C}^2}-\zeta\Theta
+\frac{\mathrm{P}_{\mathrm{r}}^{(D)}}{\mathcal{B}^{2}}+\frac{\mathcal{E}_{11}^{(D)}}{\mathcal{B}^{2}}\bigg)\bigg\}.
\end{align}
This equation demonstrates that how time produces variations in heat
energy and also guarantees that the self-gravitating system is
influenced by the thermal conductivity, temperature and relaxation
time.

Solving Eqs.\eqref{g32} and \eqref{g35} to eliminate
$\mathfrak{D}_{\mathrm{t}}\left(8\pi\varsigma-\frac{\varsigma^{(D)}}{\mathcal{AB}}-\frac{\mathcal{E}_{01}^{(D)}}{\mathcal{AB}}\right)$,
we obtain
\begin{align}\nonumber
&\mathfrak{D}_{\mathrm{t}}(\mathbb{U})\bigg(8\pi\mu+\frac{\mu^{(D)}}{\mathcal{A}^{2}}+\frac{\mathcal{E}_{00}^{(D)}}{\mathcal{A}^{2}}
+8\pi\mathrm{P}_{\mathrm{r}}-\zeta\Theta+\frac{\mathrm{P}_{\mathrm{r}}^{(D)}}{\mathcal{B}^{2}}
+\frac{\mathcal{E}_{11}^{(D)}}{\mathcal{B}^{2}}-\frac{\eta\tau}{\varrho}\bigg)=-\bigg\{\frac{\tilde{m}}{\mathcal{C}^{2}}\\\nonumber
&-\frac{\mathfrak{L}}{8\mathcal{C}^{2}}\bigg(1+\mathbb{U}^{2}+\frac{s^2}{2}\bigg)+\frac{\mathcal{C}}{1-\frac{\Psi{s^2}}{2\mathcal{C}^2}}
\bigg(8\pi\mathrm{P}_{\mathrm{r}}-\frac{s^2}{\mathcal{C}^2}-\zeta\Theta+\frac{\mathrm{P}_{\mathrm{r}}^{(D)}}{\mathcal{B}^{2}}
+\frac{\mathcal{E}_{11}^{(D)}}{\mathcal{B}^{2}}\bigg)\bigg\}\\\nonumber
&\times\bigg(8\pi\mu+\frac{\mu^{(D)}}{\mathcal{A}^{2}}+\frac{\mathcal{E}_{00}^{(D)}}{\mathcal{A}^{2}}
+8\pi\mathrm{P}_{\mathrm{r}}-\zeta\Theta+\frac{\mathrm{P}_{\mathrm{r}}^{(D)}}{\mathcal{B}^{2}}
+\frac{\mathcal{E}_{11}^{(D)}}{\mathcal{B}^{2}}-\frac{\eta\tau}{\varrho}\bigg)\bigg\{1-\frac{\eta\tau}{\varrho}\\\nonumber
&\times\bigg(8\pi\mu+\frac{\mu^{(D)}}{\mathcal{A}^{2}}+\frac{\mathcal{E}_{00}^{(D)}}{\mathcal{A}^{2}}
+8\pi\mathrm{P}_{\mathrm{r}}-\zeta\Theta+\frac{\mathrm{P}_{\mathrm{r}}^{(D)}}{\mathcal{B}^{2}}
+\frac{\mathcal{E}_{11}^{(D)}}{\mathcal{B}^{2}}\bigg)^{-1}\bigg\}+\omega^{2}\bigg\{\frac{\eta\tau\mathfrak{L}}{8\varrho\mathcal{C}^{2}}\\\nonumber
&-\frac{\mathfrak{L}}{8\mathcal{C}^{2}}\bigg(8\pi\mu+\frac{\mu^{(D)}}{\mathcal{A}^{2}}+\frac{\mathcal{E}_{00}^{(D)}}{\mathcal{A}^{2}}
+8\pi\mathrm{P}_{\mathrm{r}}-\zeta\Theta+\frac{\mathrm{P}_{\mathrm{r}}^{(D)}}{\mathcal{B}^{2}}
+\frac{\mathcal{E}_{11}^{(D)}}{\mathcal{B}^{2}}\bigg)-\frac{1}{\mathcal{C}}\\\nonumber
&\times\bigg(8\pi\mathrm{P}_{\mathrm{r}}-\frac{2s^2}{\mathcal{C}^2}+\frac{\mathrm{P}_{\mathrm{r}}^{(D)}}{\mathcal{B}^{2}}
+\frac{\mathcal{E}_{11}^{(D)}}{\mathcal{B}^{2}}-8\pi\mathrm{P}_{\phi}-\frac{\mathrm{P}_{\phi}^{(D)}}{\mathcal{C}^{2}}\bigg)\bigg\}
-\omega\bigg[-\frac{\eta\tau'}{\varrho\mathcal{B}}+\frac{1}{\mathcal{B}}\\\nonumber
&\times\bigg(8\pi\mathrm{P}_{\mathrm{r}}-\frac{s^2}{\mathcal{C}^2}-\zeta\Theta+\frac{\mathrm{P}_{\mathrm{r}}^{(D)}}{\mathcal{B}^{2}}
+\frac{\mathcal{E}_{11}^{(D)}}{\mathcal{B}^{2}}\bigg)'+\frac{1}{2}\bigg(8\pi\varsigma-\frac{\varsigma^{(D)}}{\mathcal{AB}}
-\frac{\mathcal{E}_{01}^{(D)}}{\mathcal{AB}}\bigg)\\\label{g36}
&\times\bigg\{\frac{\dot{\mathcal{B}}}{\mathcal{AB}}+\frac{\dot{\mathcal{C}}}{\mathcal{AC}}-\frac{2}{\varrho}
-\frac{\eta\tau^2}{\mathcal{A}\varrho}\bigg(\frac{\varrho}{\eta\tau^2}\bigg)^.\bigg\}\bigg],
\end{align}
whose concise form is given as
\begin{align}\nonumber
&\mathfrak{D}_{\mathrm{t}}(\mathbb{U})\bigg(8\pi\mu+\frac{\mu^{(D)}}{\mathcal{A}^{2}}+\frac{\mathcal{E}_{00}^{(D)}}{\mathcal{A}^{2}}
+8\pi\mathrm{P}_{\mathrm{r}}-\zeta\Theta+\frac{\mathrm{P}_{\mathrm{r}}^{(D)}}{\mathcal{B}^{2}}
+\frac{\mathcal{E}_{11}^{(D)}}{\mathcal{B}^{2}}\bigg)\big(1-\mathfrak{H}\big)=\mathcal{F}_{\mathrm{hyd}}
\\\nonumber
&-\mathcal{F}_{\mathrm{grav}}\big(1-\mathfrak{H}\big)+\omega^{2}\bigg\{\frac{\eta\tau\mathfrak{L}}{8\varrho\mathcal{C}^{2}}
-\frac{\mathfrak{L}}{8\mathcal{C}^{2}}\bigg(8\pi\mu+\frac{\mu^{(D)}}{\mathcal{A}^{2}}+\frac{\mathcal{E}_{00}^{(D)}}{\mathcal{A}^{2}}
+8\pi\mathrm{P}_{\mathrm{r}}-\zeta\Theta\\\label{g37}
&+\frac{\mathrm{P}_{\mathrm{r}}^{(D)}}{\mathcal{B}^{2}}
+\frac{\mathcal{E}_{11}^{(D)}}{\mathcal{B}^{2}}\bigg)-\frac{1}{\mathcal{C}}\bigg(8\pi\mathrm{P}_{\mathrm{r}}
-\frac{2s^2}{\mathcal{C}^2}+\frac{\mathrm{P}_{\mathrm{r}}^{(D)}}{\mathcal{B}^{2}}+\frac{\mathcal{E}_{11}^{(D)}}{\mathcal{B}^{2}}
-8\pi\mathrm{P}_{\phi}-\frac{\mathrm{P}_{\phi}^{(D)}}{\mathcal{C}^{2}}\bigg)\bigg\},
\end{align}
with
\begin{align}\label{g38}
\mathfrak{H}=&\frac{\eta\tau}{\varrho}\bigg(8\pi\mu+\frac{\mu^{(D)}}{\mathcal{A}^{2}}+\frac{\mathcal{E}_{00}^{(D)}}{\mathcal{A}^{2}}
+8\pi\mathrm{P}_{\mathrm{r}}-\zeta\Theta+\frac{\mathrm{P}_{\mathrm{r}}^{(D)}}{\mathcal{B}^{2}}
+\frac{\mathcal{E}_{11}^{(D)}}{\mathcal{B}^{2}}\bigg)^{-1},\\\nonumber
\mathcal{F}_{\mathrm{grav}}=&\bigg(8\pi\mu+\frac{\mu^{(D)}}{\mathcal{A}^{2}}+\frac{\mathcal{E}_{00}^{(D)}}{\mathcal{A}^{2}}
+8\pi\mathrm{P}_{\mathrm{r}}-\zeta\Theta+\frac{\mathrm{P}_{\mathrm{r}}^{(D)}}{\mathcal{B}^{2}}
+\frac{\mathcal{E}_{11}^{(D)}}{\mathcal{B}^{2}}\bigg)
\bigg\{\frac{\tilde{m}}{\mathcal{C}^{2}}-\frac{\mathfrak{L}}{8\mathcal{C}^{2}}\\\label{g39}
&\times\bigg(\mathbb{U}^{2}+1+\frac{s^2}{2}\bigg)+\frac{\mathcal{C}}{1-\frac{\Psi{s^2}}{2\mathcal{C}^2}}\bigg(8\pi\mathrm{P}_{\mathrm{r}}
-\frac{s^2}{\mathcal{C}^2}-\zeta\Theta+\frac{\mathrm{P}_{\mathrm{r}}^{(D)}}{\mathcal{B}^{2}}
+\frac{\mathcal{E}_{11}^{(D)}}{\mathcal{B}^{2}}\bigg)\bigg\},\\\nonumber
\mathcal{F}_{\mathrm{hyd}}=&-\omega\bigg[\frac{1}{\mathcal{B}}\bigg(8\pi\mathrm{P}_{\mathrm{r}}-\frac{s^2}{\mathcal{C}^2}-\zeta\Theta
+\frac{\mathrm{P}_{\mathrm{r}}^{(D)}}{\mathcal{B}^{2}}+\frac{\mathcal{E}_{11}^{(D)}}{\mathcal{B}^{2}}\bigg)'
+\frac{1}{2}\bigg\{\frac{\dot{\mathcal{B}}}{\mathcal{AB}}+\frac{\dot{\mathcal{C}}}{\mathcal{AC}}-\frac{2}{\varrho}\\\label{g40}
&-\frac{\eta\tau^2}{\mathcal{A}\varrho}\bigg(\frac{\varrho}{\eta\tau^2}\bigg)^.\bigg\}\bigg(8\pi\varsigma-\frac{\varsigma^{(D)}}{\mathcal{AB}}
-\frac{\mathcal{E}_{01}^{(D)}}{\mathcal{AB}}\bigg)-\frac{\eta\tau'}{\varrho\mathcal{B}}\bigg].
\end{align}

We can observe from Eq.\eqref{g37} that several forces involving
Newtonian ($\mathcal{F}_{\mathrm{newtn}}$), hydrodynamical
($\mathcal{F}_{\mathrm{hyd}}$) and gravitational
($\mathcal{F}_{\mathrm{grav}}$) affect the collapse rate. It is
understood that energy of the system always dissipates from its
higher to lower state in the form of radiation, convection and
conduction. If the higher phase of a celestial structure provides
energy to photons, that energy will be dissipated through
radiations. However, heat dissipates in the form of convection if
photons do not occupy all energy of the star. The cooler gases move
themselves to the hot zone in this phenomenon to attain energy,
whereas the hot gasses travel towards the upper zone and thus their
energy is radiated. Inside a system, atoms collide continuously
which results in the transfer of energy from an atom to its nearest
one, and thus energy dissipates by conduction.

The fulfillment of the equivalence principle can be observed by an
entity $(1-\mathfrak{H})$ appearing in Eq.\eqref{g37}. There exist
an inverse relation between the gravitational mass density and the
term $(\mathfrak{H})$ \big(provided in Eq.\eqref{g38}\big). This
leads to some different cases as the gravitational force of the
system is strongly affected by $1-\mathfrak{H}$. In the following,
we provide different possibilities for the collapse rate to be
increased or decreased as follows.
\begin{itemize}
\item For $\mathfrak{H}<1$, the entity $(1-\mathfrak{H})$ becomes positive which further produces negative gravitational force
(or repulsive force) as the minus sign appears in the first term on
right side of Eq.\eqref{g37}. Thus, the collapse rate will
ultimately be diminished.

\item The negative effects of $1-\mathfrak{H}$ (i.e., $\mathfrak{H}>1$) will increase the gravitational force as well as the rate
of the cylindrical collapse.

\item Moreover, the two forces (gravitational and inertial) disappear for $\mathfrak{H}=1$, hence, Eq.\eqref{g37}
is left with
\begin{align}\nonumber
&\omega^{2}\bigg\{\frac{\eta\tau\mathfrak{L}}{8\varrho\mathcal{C}^{2}}-\frac{\mathfrak{L}}{8\mathcal{C}^{2}}
\bigg(8\pi\mu+\frac{\mu^{(D)}}{\mathcal{A}^{2}}+\frac{\mathcal{E}_{00}^{(D)}}{\mathcal{A}^{2}}+8\pi\mathrm{P}_{\mathrm{r}}
-\zeta\Theta+\frac{\mathrm{P}_{\mathrm{r}}^{(D)}}{\mathcal{B}^{2}}+\frac{\mathcal{E}_{11}^{(D)}}{\mathcal{B}^{2}}\bigg)\\\nonumber
&-\frac{1}{\mathcal{C}}\bigg(8\pi\mathrm{P}_{\mathrm{r}}-\frac{2s^2}{\mathcal{C}^2}+\frac{\mathrm{P}_{\mathrm{r}}^{(D)}}{\mathcal{B}^{2}}
+\frac{\mathcal{E}_{11}^{(D)}}{\mathcal{B}^{2}}-8\pi\mathrm{P}_{\phi}
-\frac{\mathrm{P}_{\phi}^{(D)}}{\mathcal{C}^{2}}\bigg)\bigg\}=\omega\\\nonumber
&\times\bigg[\frac{1}{\mathcal{B}}\bigg(8\pi\mathrm{P}_{\mathrm{r}}-\frac{s^2}{\mathcal{C}^2}-\zeta\Theta
+\frac{\mathrm{P}_{\mathrm{r}}^{(D)}}{\mathcal{B}^{2}}+\frac{\mathcal{E}_{11}^{(D)}}{\mathcal{B}^{2}}\bigg)'
+\frac{1}{2}\bigg\{\frac{\dot{\mathcal{B}}}{\mathcal{AB}}+\frac{\dot{\mathcal{C}}}{\mathcal{AC}}-\frac{2}{\varrho}\\\label{g41}
&-\frac{\eta\tau^2}{\mathcal{A}\varrho}\bigg(\frac{\varrho}{\eta\tau^2}\bigg)^.\bigg\}\bigg(8\pi\varsigma-\frac{\varsigma^{(D)}}{\mathcal{AB}}
-\frac{\mathcal{E}_{01}^{(D)}}{\mathcal{AB}}\bigg)-\frac{\eta\tau'}{\varrho\mathcal{B}}\bigg].
\end{align}
\end{itemize}
This equation declares that the collapsing phenomenon is also
influenced by the temperature, bulk viscosity and the thermal
conductivity along with modified corrections. The expression on the
left side presents the hydrodynamical force which supports the
cylinder to maintain its equilibrium position, and consequently
results in the reduction of collapse rate.

\section{Relation between the Weyl Scalar and Effective Physical Quantities}

This section explores some substantial relations between the Weyl
scalar
($\mathbb{C}^2=\mathfrak{C}_{\mu\varphi\nu\vartheta}\mathfrak{C}^{\mu\varphi\nu\vartheta}$,
where $\mathfrak{C}_{\mu\varphi\nu\vartheta}$ represents the Weyl
tensor) and effective state variables. The linear combination of the
Ricci tensor ($\mathcal{R}_{\varphi\vartheta}$), the Ricci scalar
and the Kretchmann scalar (contraction of the Riemann tensor with
itself, i.e.,
$\mathbb{R}=\mathcal{R}_{\mu\varphi\nu\vartheta}\mathcal{R}^{\mu\varphi\nu\vartheta}$)
expresses the scalar $\mathbb{C}^2$ as \cite{35c}
\begin{align}\label{g42}
\mathbb{C}^2=-\frac{1}{3}\left(6\mathcal{R}_{\varphi\vartheta}\mathcal{R}^{\varphi\vartheta}-\mathcal{R}^2-3\mathbb{R}\right).
\end{align}
One can manipulate the scalar $\mathbb{R}$ as
\begin{align}\label{g43}
\mathbb{R}=\frac{2}{\mathcal{A}^4\mathcal{B}^4\mathcal{C}^4}\left\{2\mathcal{A}^4\big(\mathcal{R}^{1212}\big)^2
+2\mathcal{B}^4\big(\mathcal{R}^{0202}\big)^2+2\mathcal{C}^4\big(\mathcal{R}^{0101}\big)^2
-\mathcal{A}^2\mathcal{B}^2\big(\mathcal{R}^{1202}\big)^2\right\}.
\end{align}
The cylinder \eqref{g6} gives different geometric quantities such as
the Ricci scalar, non-vanishing elements of the Riemann tensor and
the Ricci tensor in relation with the Einstein tensor as
\begin{align}\nonumber
\mathcal{R}&=-2\bigg(\frac{\mathcal{G}_{22}}{\mathcal{C}^2}+\frac{\mathcal{G}_{11}}{\mathcal{B}^2}
-\frac{\mathcal{G}_{00}}{\mathcal{A}^2}\bigg),\\\nonumber
\mathcal{R}^{0202}&=\frac{\mathcal{G}_{11}}{\big(\mathcal{ABC}\big)^2},\quad
\mathcal{R}^{0101}=\frac{\mathcal{G}_{22}}{\big(\mathcal{ABC}\big)^2},\quad
\mathcal{R}^{1212}=\frac{\mathcal{G}_{00}}{\big(\mathcal{ABC}\big)^2},\\\nonumber
\mathcal{R}^{0212}&=\frac{\mathcal{G}_{01}}{\big(\mathcal{ABC}\big)^2},\quad
\mathcal{R}_{01}=\mathcal{G}_{01}, \quad
\mathcal{R}_{22}=\mathcal{C}^2\bigg(\frac{\mathcal{G}_{00}}{\mathcal{A}^2}-\frac{\mathcal{G}_{11}}{\mathcal{B}^2}\bigg),\\\nonumber
\mathcal{R}_{11}&=\mathcal{B}^2\bigg(\frac{\mathcal{G}_{00}}{\mathcal{A}^2}-\frac{\mathcal{G}_{22}}{\mathcal{C}^2}\bigg),\quad
\mathcal{R}_{00}=\mathcal{A}^2\bigg(\frac{\mathcal{G}_{11}}{\mathcal{B}^2}+\frac{\mathcal{G}_{22}}{\mathcal{C}^2}\bigg),
\end{align}
which enforce Eq.\eqref{g43} to take the following form
\begin{align}\label{g44}
\mathbb{R}=-\frac{4}{\mathcal{A}^4\mathcal{B}^4\mathcal{C}^4}\left\{4\mathcal{A}^2\mathcal{B}^2\mathcal{C}^4\mathcal{G}_{01}^2
-\mathcal{B}^4\mathcal{C}^4\mathcal{G}_{00}^2-\mathcal{A}^4\mathcal{C}^4\mathcal{G}_{11}^2-\mathcal{A}^4\mathcal{B}^4\mathcal{G}_{22}^2\right\}.
\end{align}
These preceding equations can be inserted in Eq.\eqref{g42} to
obtain the Weyl scalar as
\begin{align}\nonumber
\mathbb{C}^2&=\frac{4}{3\mathcal{A}^4\mathcal{B}^4\mathcal{C}^4}\big\{\mathcal{A}^4\mathcal{B}^4\mathcal{G}_{22}^2
+\mathcal{A}^4\mathcal{C}^4\mathcal{G}_{11}^2+\mathcal{B}^4\mathcal{C}^4\mathcal{G}_{00}^2\\\label{g45}
&+\mathcal{A}^2\mathcal{B}^2\mathcal{C}^2\big(\mathcal{C}^2\mathcal{G}_{00}\mathcal{G}_{11}
+\mathcal{B}^2\mathcal{G}_{00}\mathcal{G}_{22}-\mathcal{A}^2\mathcal{G}_{11}\mathcal{G}_{22}\big)\big\}.
\end{align}
After making use of the field equations \eqref{g8}-\eqref{g8b} in
Eq.\eqref{g45}, we have
\begin{align}\nonumber
\frac{\sqrt{3}\mathbb{C}}{2}&=\bigg[\bigg\{\frac{1}{1-\frac{\Psi{s^2}}{2\mathcal{C}^2}}\bigg(8\pi\mu+\frac{\mu^{(D)}}{\mathcal{A}^{2}}
+\frac{\mathcal{E}_{00}^{(D)}}{\mathcal{A}^{2}}+8\pi\mathrm{P}_{\mathrm{r}}+\frac{\mathrm{P}_{\mathrm{r}}^{(D)}}{\mathcal{B}^{2}}
+\frac{\mathcal{E}_{11}^{(D)}}{\mathcal{B}^{2}}-8\pi\mathrm{P}_{\phi}\\\nonumber
&-\frac{s^2}{\mathcal{C}^2}-\frac{\mathrm{P}_{\phi}^{(D)}}{\mathcal{C}^{2}}\bigg)\bigg\}^2
-\frac{1}{1-\frac{\Psi{s^2}}{2\mathcal{C}^2}}\bigg\{\bigg(8\pi\mathrm{P}_{\mathrm{r}}-\frac{s^2}{\mathcal{C}^2}-\zeta\Theta
+\frac{\mathrm{P}_{\mathrm{r}}^{(D)}}{\mathcal{B}^{2}}+\frac{\mathcal{E}_{11}^{(D)}}{\mathcal{B}^{2}}\bigg)\\\nonumber
&\bigg(8\pi\mathrm{P}_{\phi}+\frac{s^2}{\mathcal{C}^2}-\zeta\Theta
+\frac{\mathrm{P}_{\phi}^{(D)}}{\mathcal{C}^{2}}\bigg)
+\bigg(8\pi\mu+\frac{s^2}{\mathcal{C}^2}+\frac{\mu^{(D)}}{\mathcal{A}^{2}}+\frac{\mathcal{E}_{00}^{(D)}}{\mathcal{A}^{2}}\bigg)\\\label{g46}
&\times\bigg(8\pi\mathrm{P}_{\mathrm{r}}-\frac{4s^2}{\mathcal{C}^2}+2\zeta\Theta
+\frac{\mathrm{P}_{\mathrm{r}}^{(D)}}{\mathcal{B}^{2}}+\frac{\mathcal{E}_{11}^{(D)}}{\mathcal{B}^{2}}-24\pi\mathrm{P}_{\phi}
-\frac{3\mathrm{P}_{\phi}^{(D)}}{\mathcal{C}^{2}}\bigg)\bigg\}\bigg]^{\frac{1}{2}}.
\end{align}

The homogeneous energy density in the interior of any celestial body
is necessary and sufficient condition for that object to be
conformally flat. We check whether this result is valid in the
present scenario of extended gravity or not. In this regard, the
standard model \eqref{g5d} has been considered. To ensure validity
of the above result, we assume $\mathcal{R}=\mathcal{R}_0$ and treat
$f_2(\mathcal{T}=\mathcal{T}_0)$ as well as
$f_3(\mathcal{Q}=\mathcal{Q}_0)$ as constants, Eq.\eqref{g46}
ultimately produces
\begin{align}\nonumber
\frac{\sqrt{3}\mathbb{C}}{2}&=\bigg[\bigg\{\frac{8\pi}{1-\frac{\Psi{s^2}}{2\mathcal{C}^2}}\bigg(\mu+\mathrm{P}_{\mathrm{r}}-\mathrm{P}_{\phi}
-\frac{s^2}{8\pi\mathcal{C}^2}-\frac{\mathcal{C}_{0}}{16\pi}\bigg)\bigg\}^2-\frac{1}{1-\frac{\Psi{s^2}}{2\mathcal{C}^2}}\\\nonumber
&\times\bigg\{\bigg(8\pi\mathrm{P}_{\mathrm{r}}+2\zeta\Theta-24\pi\mathrm{P}_{\phi}-\mathcal{C}_{0}-\frac{4s^2}{\mathcal{C}^2}\bigg)
\bigg(8\pi\mu+\frac{s^2}{\mathcal{C}^2}-\frac{\mathcal{C}_{0}}{2}\bigg)\\\label{g47}
&+\bigg(8\pi\mathrm{P}_{\mathrm{r}}-\frac{s^2}{\mathcal{C}^2}-\zeta\Theta+\frac{\mathcal{C}_{0}}{2}\bigg)
\bigg(8\pi\mathrm{P}_{\phi}+\frac{s^2}{\mathcal{C}^2}-\zeta\Theta+\frac{\mathcal{C}_{0}}{2}\bigg)\bigg\}\bigg]^{\frac{1}{2}},
\end{align}
where
$\mathcal{C}_{0}=\Phi\sqrt{\mathcal{T}_{0}}+\Psi\mathcal{Q}_{0}$
specifies a constant quantity. Equation \eqref{g47} interprets that
the appearance of principal stresses, electric charge and the bulk
viscosity guarantee the presence of inhomogeneity in the energy
density of the matter source. This relation also demonstrates that
the tidal forces may cause the increment in inhomogeneity of the
system (during evolution) \cite{36a}. The conformally flat geometry
can only be obtained by neglecting above-mentioned factors, thus we
have from Eq.\eqref{g47} as
\begin{align}\label{g48}
\sqrt{3}\mathbb{C}&=16\pi\bigg(\mu-\frac{\mathcal{C}_{0}}{16\pi}\bigg)
\quad \Rightarrow \quad \sqrt{3}\mathbb{C}'=16\pi\mu'.
\end{align}
This equation ensures that homogeneity in the energy density (i.e.,
$\mu'=0$) results in the conformally flat spacetime (as regular axis
condition gives $\mathbb{C}=0$) and contrariwise, and hence we
obtain the required condition.

\section{Final Remarks}

The constituents that make up our cosmos contain plenty of
astronomical bodies. An appealing phenomenon in this regard is the
gravitational collapse which leads to the structural formation of
those celestial objects. The study of gravitational waves through
many observations (i.e., laser interferometric detectors like LIGO,
VIRGO, GEO and TAMA) has prompted several astronomers to analyze the
collapsing rate of self-gravitating geometries in $\mathbb{GR}$ and
other extended theories \cite{44}. This article formulates the
dynamical description of the self-gravitating cylindrical matter
configuration influenced by electromagnetic field to study the
evolutionary changes gradually produced by different physical
factors within the system in
$f(\mathcal{R},\mathcal{T},\mathcal{R}_{\phi\psi}\mathcal{T}^{\phi\psi})$
gravitational theory. We have assumed that the interior geometry
\eqref{g6} involves the effects of heat dissipation, expansion
scalar and principal pressures in three different directions (such
as $\mathrm{P}_{\mathrm{r}},~\mathrm{P}_{\phi}$ and
$\mathrm{P}_{\mathrm{z}}$). The non-zero components of the Bianchi
identities have been formulated through Misner-Sharp formalism. We
have calculated the corresponding C-energy and its proper radial as
well as temporal derivatives to examine the variations in the total
energy of the system.

We have identified some forces (such as Newtonian, hydrodynamical
and gravitational) by constructing the transport equation and then
studied the effects of charge and modified corrections on the
collapse rate by coupling these forces with the dynamical equations.
We have defined the entity $\mathfrak{H}$ which is found to be
directly related with thermal conductivity as well as temperature
and in inverse relation with the gravitational mass density. The
results produced by this extended gravity are summarized in the
following.
\begin{itemize}
\item The positive effect of $f(\mathcal{R},\mathcal{T},\mathcal{R}_{\phi\psi}\mathcal{T}^{\phi\psi})$ corrections
results in the entity $\mathfrak{H}$ smaller than that in
$\mathbb{GR}$. This consequently increases the term
$(1-\mathfrak{H})$ as well as the gravitational force which supports
the star to collapse. Meanwhile, this force is appeared with minus
sign that ultimately assures decrement in the collapse rate.

\item The increment in cylindrical collapse rate can be observed for the negative effect of correction terms which
may decrease due to the charge.

\item If the curvature terms due to the extended theory have opposite signs, then we cannot declare the rate of collapse
to be either increased or decreased.

\item The energy of the considered matter source reduces as a whole
due to the involvement of electromagnetic field, which reveals the
heat to be dissipated in outward direction.
\end{itemize}

We have also established an appropriate relationship between the
Weyl scalar and state variables such as inhomogeneous energy density
and principal pressures to check the fulfillment of the spacetime to
be conformally flat. After imposing some constraints on the modified
model \eqref{g5d} along with the regular axis condition, it is
observed that the density homogeneity and conformal flatness of the
considered matter source imply each other. It is worth mentioning
that the fluid becomes more inhomogeneous during the evolutionary
process in this case due to the tidal forces. All these results can
be retrieved in $\mathbb{GR}$ by substituting $\Phi=0=\Psi$.

\section*{Appendix A}

\renewcommand{\theequation}{A\arabic{equation}}
\setcounter{equation}{0} The
$f(\mathcal{R},\mathcal{T},\mathcal{Q})$ corrections in the field
equations \eqref{g8}-\eqref{g8d} are
\begin{align}\nonumber
\mu^{(D)}&=-\frac{\mathcal{A}^2\big(\Phi\sqrt{\mathcal{T}}+\Psi\mathcal{Q}\big)}{2}+\Psi\bigg\{\mu\bigg(\frac{4\dot{\mathcal{A}}^2}{\mathcal{A}^2}
-\frac{\mathcal{A}'^2}{\mathcal{B}^2}+\frac{\mathcal{AA}''}{\mathcal{B}^2}+\frac{3\dot{\mathcal{A}}\dot{\mathcal{B}}}{\mathcal{AB}}
-\frac{\mathcal{AA}'\mathcal{B}'}{\mathcal{B}^3}\\\nonumber
&+\frac{3\dot{\mathcal{A}}\dot{\mathcal{C}}}{\mathcal{AC}}+\frac{\mathcal{AA}'\mathcal{C}'}{\mathcal{B}^2\mathcal{C}}
-\frac{2\ddot{\mathcal{C}}}{\mathcal{C}}-\frac{2\ddot{\mathcal{B}}}{\mathcal{B}}\bigg)+\dot{\mu}\bigg(\frac{\dot{\mathcal{C}}}{2\mathcal{C}}
+\frac{\dot{\mathcal{B}}}{2\mathcal{B}}\bigg)-\mu'\bigg(\frac{2\mathcal{AA}'}{\mathcal{B}^2}
-\frac{\mathcal{A}^2\mathcal{B}'}{2\mathcal{B}^3}\\\nonumber
&+\frac{\mathcal{A}^2\mathcal{C}'}{2\mathcal{B}^2\mathcal{C}}\bigg)-\frac{\mu''\mathcal{A}^2}{2\mathcal{B}^2}
+\mathrm{P}_\mathrm{r}\bigg(\frac{4\mathcal{A}^2\mathcal{B}'^2}{\mathcal{B}^4}-\frac{\mathcal{A}^2\mathcal{B}''}{\mathcal{B}^3}
+\frac{\dot{\mathcal{B}}^2}{\mathcal{B}^2}\bigg)-\frac{\dot{\mathrm{P}}_\mathrm{r}\dot{\mathcal{B}}}{2\mathcal{B}}
-\frac{5\mathrm{P}'_\mathrm{r}\mathcal{A}^2\mathcal{B}'}{2\mathcal{B}^3}\\\nonumber
&+\frac{\mathrm{P}''_\mathrm{r}\mathcal{A}^2}{2\mathcal{B}^2}-\mathrm{P}_{\phi}\bigg(\frac{\dot{\mathcal{C}}^2}{\mathcal{C}^2}
-\frac{\mathcal{A}^2\mathcal{C}'^2}{\mathcal{B}^2\mathcal{C}^2}\bigg)-\frac{\dot{\mathrm{P}}_{\phi}\dot{\mathcal{C}}}{2\mathcal{C}}
+\frac{\mathrm{P}'_{\phi}\mathcal{A}^2\mathcal{C}'}{2\mathcal{B}^2\mathcal{C}}-\varsigma\bigg(\frac{2\mathcal{A}\dot{\mathcal{C}}'}{\mathcal{BC}}
-\frac{2\dot{\mathcal{A}}\mathcal{B}'}{\mathcal{B}^2}\\\nonumber
&+\frac{2\dot{\mathcal{A}}'}{\mathcal{B}}-\frac{4\dot{\mathcal{A}}\mathcal{A}'}{\mathcal{AB}}-\frac{2\mathcal{A}'\dot{\mathcal{B}}}{\mathcal{B}^2}
-\frac{2\mathcal{A}\dot{\mathcal{B}}\mathcal{C}'}{\mathcal{B}^2\mathcal{C}}-\frac{2\mathcal{A}'\dot{\mathcal{C}}}{\mathcal{BC}}
-\frac{2\mathcal{A}'\dot{\mathcal{B}}}{\mathcal{B}^2}\bigg)-\frac{2\dot{\varsigma}\mathcal{A}'}{\mathcal{B}}
-\frac{2\varsigma'\dot{\mathcal{A}}}{\mathcal{B}}\bigg\},\\\nonumber
\mathrm{P}_{\mathrm{r}}^{(D)}&=\frac{\mathcal{B}^2}{2}\bigg\{\bigg(\frac{\Phi}{\sqrt{\mathcal{T}}}+\Psi\mathcal{R}\bigg)\mathrm{P}_{\mathrm{r}}
+\Phi\sqrt{\mathcal{T}}+\Psi\mathcal{Q}+\frac{\Phi\mu}{\sqrt{\mathcal{T}}}\bigg\}
+\Psi\bigg\{\mu\bigg(\frac{\ddot{\mathcal{A}}\mathcal{B}^2}{\mathcal{A}^3}-\frac{\mathcal{A}'^2}{\mathcal{A}^2}\\\nonumber
&-\frac{4\dot{\mathcal{A}}^2B^2}{\mathcal{A}^4}\bigg)+\frac{5\dot{\mu}\dot{\mathcal{A}}B^2}{2\mathcal{A}^3}
+\frac{\mu'\mathcal{A}'}{2\mathcal{A}}-\frac{\ddot{\mu}\mathcal{B}^2}{2\mathcal{A}^2}
+\mathrm{P}_\mathrm{r}\bigg(\frac{\mathcal{B}\dot{\mathcal{A}}\dot{\mathcal{B}}}{\mathcal{A}^3}-\frac{3\mathcal{A}'\mathcal{B}'}{\mathcal{AB}}
-\frac{\mathcal{B}\dot{\mathcal{B}}\dot{\mathcal{C}}}{\mathcal{A}^2C}\\\nonumber
&+\frac{\dot{\mathcal{B}}^2}{\mathcal{A}^2}-\frac{4\mathcal{B}'^2}{\mathcal{B}^2}-\frac{\mathcal{B}\ddot{\mathcal{B}}}{\mathcal{A}^2}
-\frac{3\mathcal{B}'\mathcal{C}'}{\mathcal{BC}}+\frac{2\mathcal{A}''}{\mathcal{A}}+\frac{2\mathcal{C}''}{\mathcal{C}}\bigg)
+\dot{\mathrm{P}}_\mathrm{r}\bigg(\frac{\mathcal{B}^2\dot{\mathcal{C}}}{2\mathcal{A}^2\mathcal{C}}
-\frac{\mathcal{B}^2\dot{\mathcal{A}}}{2\mathcal{A}^3}\\\nonumber
&+\frac{2\mathcal{B}\dot{\mathcal{B}}}{\mathcal{A}^2}\bigg)-\mathrm{P}'_\mathrm{r}\bigg(\frac{\mathcal{A}'}{2\mathcal{A}}
+\frac{\mathcal{C}'}{2\mathcal{C}}\bigg)+\frac{\ddot{\mathrm{P}}_\mathrm{r}\mathcal{B}^2}{2\mathcal{A}^2}
-\mathrm{P}_{\phi}\bigg(\frac{B^2\dot{\mathcal{C}}^2}{A^2\mathcal{C}^2}-\frac{\mathcal{C}'^2}{\mathcal{C}^2}\bigg)
-\frac{\dot{\mathrm{P}}_{\phi}B^2\dot{\mathcal{C}}}{2A^2\mathcal{C}}\\\nonumber
&-\frac{\mathrm{P}'_{\phi}\mathcal{C}'}{2\mathcal{C}}+\varsigma\bigg(\frac{2\dot{\mathcal{B}}'}{\mathcal{A}}
-\frac{2\dot{\mathcal{A}}\mathcal{B}'}{\mathcal{A}^2}-\frac{4\mathcal{A}'\dot{\mathcal{B}}}{\mathcal{A}^2}
-\frac{4\dot{\mathcal{B}}\mathcal{B}'}{\mathcal{AB}}+\frac{2\mathcal{B}\dot{\mathcal{C}}'}{\mathcal{AC}}
-\frac{2\dot{\mathcal{B}}\mathcal{C}'}{\mathcal{AC}}-\frac{2\mathcal{B}\mathcal{A}'\dot{\mathcal{C}}}{\mathcal{A}^2\mathcal{C}}\bigg)\\\nonumber
&+\frac{2\dot{\varsigma}\mathcal{B}'}{\mathcal{A}}+\frac{2\varsigma'\dot{\mathcal{B}}}{\mathcal{A}}\bigg\},\\\nonumber
\mathrm{P}_{\phi}^{(D)}&=\frac{\mathcal{C}^2}{2}\bigg\{\bigg(\frac{\Phi}{\sqrt{\mathcal{T}}}+\Psi\mathcal{R}\bigg)\mathrm{P}_{\phi}
+\Phi\sqrt{\mathcal{T}}+\Psi\mathcal{Q}+\frac{\Phi\mu}{\sqrt{\mathcal{T}}}\bigg\}+\Psi\bigg\{\mu\bigg(\frac{\ddot{\mathcal{A}}\mathcal{C}^2}
{\mathcal{A}^3}-\frac{\mathcal{A}'^2\mathcal{C}^2}{\mathcal{A}^2\mathcal{B}^2}\\\nonumber
&-\frac{4\dot{\mathcal{A}}^2\mathcal{C}^2}{\mathcal{A}^4}\bigg)+\frac{5\dot{\mu}\dot{\mathcal{A}}\mathcal{C}^2}{2\mathcal{A}^3}
+\frac{\mu'\mathcal{A}'\mathcal{C}^2}{2\mathcal{A}\mathcal{B}^2}-\frac{\ddot{\mu}\mathcal{C}^2}{2\mathcal{A}^2}
+\mathrm{P}_\mathrm{r}\bigg(\frac{\mathcal{B}''\mathcal{C}^2}{\mathcal{B}^3}-\frac{4\mathcal{B}'^2\mathcal{C}^2}{\mathcal{B}^4}
-\frac{\dot{\mathcal{B}}^2\mathcal{C}^2}{\mathcal{A}^2\mathcal{B}^2}\bigg)\\\nonumber
&+\frac{\dot{\mathrm{P}}_\mathrm{r}\mathcal{C}^2\dot{\mathcal{B}}}{2\mathcal{A}^2\mathcal{B}}
+\frac{5\mathrm{P}'_\mathrm{r}\mathcal{C}^2\mathcal{B}'}{2\mathcal{B}^3}-\frac{{\mathrm{P}}''_\mathrm{r}\mathcal{C}^2}{2\mathcal{B}^2}
+\mathrm{P}_{\phi}\bigg(\frac{\dot{\mathcal{C}}^2}{\mathcal{A}^2}-\frac{\mathcal{C}\ddot{\mathcal{C}}}{\mathcal{A}^2}
+\frac{\mathcal{C}\dot{\mathcal{A}}\dot{\mathcal{C}}}{\mathcal{A}^3}-\frac{\mathcal{C}\mathcal{B}'\mathcal{C}'}{\mathcal{B}^3}
-\frac{\mathcal{C}\dot{\mathcal{B}}\dot{\mathcal{C}}}{\mathcal{A}^2\mathcal{B}}\\\nonumber
&-\frac{\mathcal{C}'^2}{\mathcal{B}^2}-\frac{\mathcal{C}\mathcal{A}'\mathcal{C}'}{\mathcal{AB}^2}+\frac{\mathcal{C}\mathcal{C}''}
{\mathcal{B}^2}\bigg)+\dot{\mathrm{P}}_{\phi}\bigg(\frac{\mathcal{C}^2\dot{\mathcal{B}}}{2\mathcal{A}^2\mathcal{B}}
-\frac{\mathcal{C}^2\dot{\mathcal{A}}}{2\mathcal{A}^3}+\frac{2\mathcal{C}\dot{\mathcal{C}}}{\mathcal{A}^2}\bigg)
-\mathrm{P}'_{\phi}\bigg(\frac{\mathcal{C}^2\mathcal{A}'}{2\mathcal{AB}^2}\\\nonumber
&-\frac{\mathcal{C}^2\mathcal{B}'}{2\mathcal{B}^3}+\frac{2\mathcal{CC}'}{\mathcal{B}^2}\bigg)
+\frac{\ddot{\mathrm{P}}_{\phi}\mathcal{C}^2}{2\mathcal{A}^2}-\frac{\mathrm{P}''_{\phi}\mathcal{C}^2}{2\mathcal{B}^2}-\varsigma\bigg(
\frac{3\mathcal{C}^2\dot{\mathcal{A}}\mathcal{A}'}{\mathcal{A}^3\mathcal{B}}+\frac{3\mathcal{C}^2\dot{\mathcal{B}}\mathcal{B}'}{\mathcal{AB}^3}
+\frac{3\mathcal{C}^2\mathcal{A}'\dot{\mathcal{B}}}{\mathcal{A}^2\mathcal{B}^2}\\\nonumber
&-\frac{\mathcal{C}^2\dot{\mathcal{A}}'}{\mathcal{A}^2\mathcal{B}}-\frac{\mathcal{C}^2\dot{\mathcal{B}}'}{\mathcal{AB}^2}
+\frac{\mathcal{C}^2\dot{\mathcal{A}}\mathcal{B}'}{\mathcal{A}^2\mathcal{B}^2}\bigg)
+\dot{\varsigma}\bigg(\frac{2\mathcal{C}^2\mathcal{A}'}{\mathcal{A}^2\mathcal{B}}+\frac{\mathcal{C}^2\mathcal{B}'}{\mathcal{AB}^2}\bigg)
+\varsigma'\bigg(\frac{\mathcal{C}^2\dot{\mathcal{A}}}{\mathcal{A}^2\mathcal{B}}
+\frac{2\mathcal{C}^2\dot{\mathcal{B}}}{\mathcal{AB}^2}\bigg)\\\nonumber
&-\frac{\dot{\varsigma}'\mathcal{C}^2}{\mathcal{AB}}\bigg\},\\\nonumber
\mathrm{P}_{\mathrm{z}}^{(D)}&=\frac{1}{2}\bigg\{\bigg(\frac{\Phi}{\sqrt{\mathcal{T}}}+\Psi\mathcal{R}\bigg)\mathrm{P}_{\phi}
+\Phi\sqrt{\mathcal{T}}+\Psi\mathcal{Q}+\frac{\Phi\mu}{\sqrt{\mathcal{T}}}\bigg\}+\Psi\bigg\{\mu\bigg(\frac{\ddot{\mathcal{A}}}
{\mathcal{A}^3}-\frac{\mathcal{A}'^2}{\mathcal{A}^2\mathcal{B}^2}\\\nonumber
&-\frac{4\dot{\mathcal{A}}^2}{\mathcal{A}^4}\bigg)+\frac{5\dot{\mu}\dot{\mathcal{A}}}{2\mathcal{A}^3}
+\frac{\mu'\mathcal{A}'}{2\mathcal{A}\mathcal{B}^2}-\frac{\ddot{\mu}}{2\mathcal{A}^2}
+\mathrm{P}_\mathrm{r}\bigg(\frac{\mathcal{B}''}{\mathcal{B}^3}-\frac{4\mathcal{B}'^2}{\mathcal{B}^4}
-\frac{\dot{\mathcal{B}}^2}{\mathcal{A}^2\mathcal{B}^2}\bigg)
+\frac{\dot{\mathrm{P}}_\mathrm{r}\dot{\mathcal{B}}}{2\mathcal{A}^2\mathcal{B}}\\\nonumber
&+\frac{5\mathrm{P}'_\mathrm{r}\mathcal{B}'}{2\mathcal{B}^3}-\frac{{\mathrm{P}}''_\mathrm{r}}{2\mathcal{B}^2}
+\mathrm{P}_{\phi}\bigg(\frac{\mathcal{C}'^2}{\mathcal{B}^2\mathcal{C}^2}-\frac{\dot{\mathcal{C}}^2}{\mathcal{A}^2\mathcal{C}^2}\bigg)
+\frac{\dot{\mathrm{P}}_{\phi}\dot{\mathcal{C}}}{2\mathcal{A}^2\mathcal{C}}-\frac{\mathrm{P}'_{\phi}\mathcal{C}'}{2\mathcal{B}^2\mathcal{C}}
+\dot{\mathrm{P}}_{\mathrm{z}}\bigg(\frac{\dot{\mathcal{B}}}{2\mathcal{A}^2\mathcal{B}}\\\nonumber
&-\frac{\dot{\mathcal{A}}}{2\mathcal{A}^3}+\frac{\dot{\mathcal{C}}}{2\mathcal{A}^2\mathcal{C}}\bigg)
+\mathrm{P}'_{\mathrm{z}}\bigg(\frac{\mathcal{B}'}{2\mathcal{B}^3}-\frac{A'}{2\mathcal{AB}^2}-\frac{\mathcal{C}'}{2\mathcal{B}^2\mathcal{C}}\bigg)
+\frac{\ddot{\mathrm{P}}_{\mathrm{z}}}{2\mathcal{A}^2}-\frac{\mathrm{P}''_{\mathrm{z}}}{2\mathcal{B}^2}\\\nonumber
&+\varsigma\bigg(\frac{\dot{\mathcal{A}}'}{\mathcal{A}^2\mathcal{B}}-\frac{3\dot{\mathcal{A}}\mathcal{A}'}{\mathcal{A}^3\mathcal{B}}
-\frac{\dot{\mathcal{A}}\mathcal{B}'}{\mathcal{A}^2\mathcal{B}^2}+\frac{\dot{\mathcal{B}}'}{\mathcal{AB}^2}
-\frac{3\mathcal{A}'\dot{\mathcal{B}}}{\mathcal{A}^2\mathcal{B}^2}-\frac{3\dot{\mathcal{B}}\mathcal{B}'}{\mathcal{AB}^3}\bigg)
+\dot{\varsigma}\bigg(\frac{2\mathcal{A}'}{\mathcal{A}^2\mathcal{B}}\\\nonumber
&+\frac{\mathcal{B}'}{\mathcal{AB}^2}\bigg)+\varsigma'\bigg(\frac{\dot{\mathcal{A}}}{\mathcal{A}^2\mathcal{B}}
+\frac{2\dot{\mathcal{B}}}{\mathcal{AB}^2}\bigg)-\frac{\dot{\varsigma}'\mathcal{C}^2}{\mathcal{AB}}\bigg\},\\\nonumber
\varsigma^{(D)}&=-\frac{\varsigma\mathcal{AB}}{2}\bigg(\frac{\Phi}{\sqrt{\mathcal{T}}}+\Psi\mathcal{R}\bigg)
+\Psi\bigg\{\mu\bigg(\frac{\mathcal{A}'\dot{\mathcal{C}}}{\mathcal{AC}}-\frac{\dot{\mathcal{C}}'}{\mathcal{C}}
+\frac{\dot{\mathcal{B}}C'}{\mathcal{BC}}\bigg)+\frac{\dot{\mu}\mathcal{A}'}{2\mathcal{A}}+\frac{\mu'\dot{\mathcal{A}}}{2\mathcal{B}}\\\nonumber
&-\frac{\dot{\mu}'}{2}+\mathrm{P}_\mathrm{r}\bigg(\frac{\dot{\mathcal{C}}'}{\mathcal{C}}-\frac{A'\dot{\mathcal{C}}}{\mathcal{AC}}
-\frac{\dot{\mathcal{B}}C'}{\mathcal{BC}}\bigg)-\frac{\dot{\mathrm{P}}_\mathrm{r}\mathcal{A}'}{2\mathcal{A}}
-\frac{\mathrm{P}'_\mathrm{r}\dot{\mathcal{B}}}{2\mathcal{B}}+\frac{\dot{\mathrm{P}}'_\mathrm{r}}{2}
+\varsigma\bigg(\frac{2\ddot{\mathcal{B}}}{\mathcal{A}}-\frac{\ddot{\mathcal{A}}\mathcal{B}}{\mathcal{A}^2}\\\nonumber
&-\frac{4\dot{\mathcal{A}}\dot{\mathcal{B}}}{\mathcal{A}^2}+\frac{2\dot{\mathcal{A}}^2\mathcal{B}}{\mathcal{A}^3}
+\frac{\mathcal{A}'^2}{\mathcal{AB}}+\frac{4\mathcal{A}'\mathcal{B}'}{\mathcal{B}^2}
-\frac{2\mathcal{A}''}{\mathcal{B}}+\frac{\mathcal{AB}''}{\mathcal{B}^2}-\frac{\dot{\mathcal{B}}^2}{\mathcal{AB}}
+\frac{\mathcal{B}\ddot{\mathcal{C}}}{\mathcal{AC}}-\frac{\mathcal{AC}''}{\mathcal{BC}}\\\nonumber
&-\frac{2\mathcal{AB}'^2}{\mathcal{B}^3}-\frac{3\mathcal{B}\dot{\mathcal{A}}\dot{\mathcal{C}}}{2\mathcal{A}^2\mathcal{C}}
+\frac{\dot{\mathcal{B}}\dot{\mathcal{C}}}{2\mathcal{AC}}-\frac{\mathcal{A}'\mathcal{C}'}{2\mathcal{BC}}
+\frac{3\mathcal{AB}'\mathcal{C}'}{2\mathcal{B}^2\mathcal{C}}\bigg)
-\dot{\varsigma}\bigg(\frac{2\dot{\mathcal{A}}\mathcal{B}}{\mathcal{A}^2}+\frac{\mathcal{B}\dot{\mathcal{C}}}{2\mathcal{AC}}\bigg)\\\nonumber
&+\varsigma'\bigg(\frac{2\mathcal{A}\mathcal{B}'}{\mathcal{B}^2}+\frac{\mathcal{A}\mathcal{C}'}{2\mathcal{BC}}\bigg)\bigg\},\\\nonumber
\mathcal{E}_{00}^{(D)}&=\frac{\Psi{s}^2}{\mathcal{AB}^3\mathcal{C}^3}\bigg(\mathcal{A}^3\mathcal{BC}''-\mathcal{AB}^2\mathcal{C}\ddot{\mathcal{B}}
+\mathcal{A}^2\mathcal{BCA}''-\mathcal{A}^3\mathcal{B}'\mathcal{C}'-\mathcal{AB}^2\dot{\mathcal{B}}\dot{\mathcal{C}}\\\nonumber
&+\mathcal{B}^2\mathcal{C}\dot{\mathcal{A}}\dot{\mathcal{B}}-\mathcal{A}^2\mathcal{CA}'\mathcal{B}'\bigg)
-\frac{\Phi{s}^2\mathcal{A}^2}{2\sqrt{\mathcal{T}}\mathcal{C}^2},\\\nonumber
\mathcal{E}_{11}^{(D)}&=\frac{\Psi{s}^2}{\mathcal{A}^3\mathcal{BC}^3}\bigg(\mathcal{AB}^3\ddot{\mathcal{C}}+\mathcal{AB}^2\mathcal{C}
\ddot{\mathcal{B}}-\mathcal{A}^2\mathcal{BCA}''-\mathcal{A}^2\mathcal{BA}'\mathcal{C}'-\mathcal{B}^3\dot{\mathcal{A}}\dot{\mathcal{C}}\\\nonumber
&-\mathcal{B}^2\mathcal{C}\dot{\mathcal{A}}\dot{\mathcal{B}}+\mathcal{A}^2\mathcal{CA}'\mathcal{B}'\bigg)
-\frac{\Phi{s}^2\mathcal{B}^2}{2\sqrt{\mathcal{T}}\mathcal{C}^2},\\\nonumber
\mathcal{E}_{01}^{(D)}&=\frac{\Psi{s}^2\mathcal{AB}}{\mathcal{C}^3}\bigg(\mathcal{A}\dot{\mathcal{B}}\mathcal{C}'
+\mathcal{B}\mathcal{A}'\dot{\mathcal{C}}-\mathcal{AB}\dot{\mathcal{C}}'\bigg).
\end{align}

\end{document}